\documentclass[twocolumn,showpacs,amsmath,amssymb,floatfix]{revtex4}

\usepackage{graphicx}
\begin{document}

\hyphenation{}
\def\W{{W^{ij\sigma}_{kl\sigma'}}}
\def\i{{ijkl\sigma\sigma'}}

\title{ Symmetry projection schemes for Gaussian Monte Carlo methods }
\author{F. F. Assaad}
\affiliation{ Institut f\"ur theoretische Physik und Astrophysik,
Universit\"at W\"urzburg, Am Hubland D-97074 W\"urzburg }
\author{P. Werner, P. Corboz,  E. Gull  and M. Troyer }
\affiliation{ Institut f\"ur theoretische Physik, ETH H\"onggerberg, CH-8093 Z\"urich, Switzerland}

\begin{abstract}
A novel sign-free Monte Carlo method for the Hubbard model has recently been proposed by Corney 
and Drummond.  High precision 
measurements on small clusters show that ground state correlation functions are not correctly reproduced. 
We argue that the origin of this mismatch lies in the fact that the low temperature density matrix 
does not have the symmetries of the Hamiltonian.  
Here we show that supplementing the algorithm with symmetry projection schemes provides
reliable and accurate estimates of ground state properties.  
\end{abstract}


\maketitle

\section{Introduction}
The understanding of low-temperature properties of doped Mott insulators is a central challenge 
in solid state physics. From the numerical point of view, this problem has remained elusive 
due to the sign problem.  Recently Corney and Drummond \cite{Corney04} have proposed  a stochastic
method in which the sign-problem does not explicitly occur.  They show that the density matrix of an 
arbitrary model Hamiltonian may be expressed as a positive 
sum over Gaussian operators. The imaginary time propagation of the density matrix  boils down to 
a Fokker-Planck equation governing the time evolution of the probability distribution in the 
space of  Gaussian operators. 
One can then solve the Fokker-Planck equation by integrating  numerically the 
corresponding stochastic differential equation (SDE). For the Hubbard model on arbitrary lattice 
topologies and at arbitrary band fillings, the SDE has real stochastic and drift forces thereby 
leading to no explicit sign problem.  

The aim of this article is to test the precision of the method with respect to ground state 
properties.  We  will see -- on the basis of simple examples --  that the low temperature density matrix
obtained by solving the SDE numerically does not have  the symmetry of the Hamiltonian thereby 
producing  biased ground state properties. The problem stems from the fact that 
a single Gaussian operator breaks spin, lattice as well as translation symmetries.  Since the 
weighted summation over the  Gaussian operators produces the density matrix, the summation has to 
restore the symmetries of the problem at hand.  However the summation is carried out stochastically, 
and it is a-priori not clear  that the sampling is efficient enough to restore  symmetries. 
At {\it high} temperatures  this poses 
no further problems but as the temperature is lowered symmetry 
restoration fails.  We show that one can solve this problem  by projecting  the 
density matrix onto the symmetry sector of the ground state. 

The organization of this paper is as follows.  In the next section, we  briefly review the 
formulation of the Gaussian Quantum Monte Carlo approach (GQMC)  applied to the Hubbard model and in Appendix~\ref{diffusion_gauge} present the generalization to four-fermion terms arising from Coulomb interactions.
We will highlight the implicit assumptions used to derive the  Fokker-Planck equation, 
absence of boundary terms,  and discuss  in Appendix \ref{Stochastic_gauges}  stochastic 
gauges which  serve as  a means to suppress them. 
In section  \ref{Symm_proj.sec} we show in detail how to implement the symmetry projections 
and demonstrate their  efficiency in section \ref{Accuracy_tests}.  Finally we draw conclusions.

\section{The Gaussian Quantum Monte Carlo Method for the Hubbard model}

In this section we  summarize the results of Ref.~\cite{Corney04}. 
Although the GQMC is general and can be generalized to arbitrary Coulomb interactions as discussed in Appendix~\ref{diffusion_gauge}, we will concentrate here on the  Hubbard model:
\begin{equation}
\label{Hubbard}
       \hat H = \hat{{\pmb c}}^{\dagger}  {\pmb T } \hat{ { \pmb c } }  - 
       \frac{U}{2} \sum_{\vec{i}}  \left(  \hat{ {\pmb c} } _{\vec{i}}^{\dagger} {\pmb \sigma}
           \hat { { \pmb c } }_{\vec{i}} \right)^2,
\end{equation} 
where $  \hat{{\bf c}}^{\dagger} = \left( \hat{c}^{\dagger}_1, \cdots, \hat{c}^{\dagger}_{N_s} \right) $. 
$\hat{c}^{\dagger}_x$ creates a fermion with quantum numbers $x = ( \vec{i}, \sigma) $ where  $\vec{i}$ 
denotes the lattice site and $\sigma$ the $z$-component of spin. Hence $x$ runs from $1 \cdots N_s \equiv 
2 N$, $N$ being the number of lattice site.
${\pmb T }$ is the hopping matrix. It is diagonal in spin indices and takes the value 
$-t$  ($-t'$) for next neighbors (next nearest neighbors). 
Finally $ {\pmb \sigma } $ denotes  a Pauli spin matrix and  
$\hat{{\pmb c}}_{\vec{i}}^{\dagger}  = \left( \hat{c}^{\dagger}_{\vec{i},\uparrow},  
\hat{c}^{\dagger}_{\vec{i},\downarrow} \right) $.
Setting $ {\pmb \sigma} = {\pmb 1 } $ yields the attractive Hubbard model whereas setting
$ {\pmb \sigma}$ to $ {\pmb \sigma}^{x} $ or $ {\pmb \sigma}^{z} $ the repulsive 
case \footnote{ In principle we could also set $ {\pmb \sigma }   = {\pmb \sigma }^{y} $.  However 
since ${\pmb \sigma }^{y}$ is purely imaginary this  leads to complex drift and stochastic forces 
and in turn to negative weights.}.

Corney and Drummond propose to expand the density matrix in terms of Gaussian operators: 
\begin{equation}
\hat{\Lambda}( {\bf n} )  = \det( {\bf 1} - {\bf n}  ) 
    :e^{ - \hat{{\bf c}}^{\dagger} \left(  {\bf 2} + \left( {\bf n}^T - {\bf 1} \right)^{-1} \right) 
           \hat{{\bf c}}   } : 
\end{equation}
with  ${\bf n} $ an $N_s \times N_s$ real matrix.  The Gaussian operators are normalized, 
$ {\rm Tr} \left[ \hat{\Lambda}( {\bf n} ) \right] = 1 $ and obey Wick's theorem such that
\begin{eqnarray}
         & & {\rm Tr} \left[ \hat{\Lambda}( {\bf n} )  \hat{c}^{\dagger}_x \hat{c}_y \right] 
= {\pmb n}_{x,y},  \nonumber \\ 
         & & {\rm Tr} \left[ \hat{\Lambda}( {\bf n} )  
       \hat{c}^{\dagger}_x \hat{c}_y   \hat{c}^{\dagger}_w \hat{c}_z  \right]  = 
       {\pmb n}_{x,y} {\pmb n}_{w,z} +   {\pmb n}_{x,z} \left( {\pmb 1} -  {\pmb n} \right)_{w,y}. 
\end{eqnarray}

The major result of Ref.~\cite{Corney04} is that one can expand the density matrix in terms of a positive sum 
of Gaussian operators: 
\begin{equation}
\label{positivity}
        \hat{\rho}(\tau) = \sum_{i} P_i(\tau) \hat{\Lambda}({\bf n}_i),  \; \; \; \;   P_i \geq 0.
\end{equation}
Clearly $ {\rm Tr} \left[ \hat{\rho}(\tau)  \right] \equiv   \sum_{i} P_i(\tau) $ grows exponentially 
with $\tau$. One can account for  this exponential growth by attaching a weight factor to the  
Gaussian operators thereby obtaining: 
\begin{eqnarray}
\label{Den_matrix}
 \hat{\rho}(\tau) & = & 
 \int {\rm d} \underline{\pmb \lambda} P(\underline{\pmb \lambda},\tau) 
 \hat{\Lambda}(\underline{\pmb \lambda}),
 \end{eqnarray}
with $ \underline{\pmb \lambda} =  (\Omega,{\pmb n})$,   
$ \hat{\Lambda}(\underline{\pmb \lambda}) = \Omega \hat{\Lambda}({\pmb n})$ and  
$ \int {\rm d} \underline{\pmb \lambda} 
P(\underline{\pmb \lambda},\tau) = 1$.

The aim is now to formulate a stochastic process which samples the probability distribution
$P(\underline{\pmb \lambda},\tau)$ in the space of Gaussian operators. To this end one considers 
the imaginary time evolution of the density matrix
\begin{equation}
        \frac {\text{d} }{\text{d} \tau}  \hat{\rho}(\tau)  =          
- \frac{1}{2} \left[\hat{H}, \hat{\rho}(\tau)  \right]_{+},
\end{equation}
so that in conjunction with Eq.~(\ref{Den_matrix})  we are left with the evaluation 
of the anti-commutator  $-\frac{1}{2} \left[\hat{H}, \hat{\Lambda} (\underline{\pmb \lambda}) \right]_{+} $. 
The anti-commutator can  be transformed into a differential form acting on 
$\hat{\Lambda} (\underline{\pmb \lambda})$: 
\begin{eqnarray}
\label{Anticomm}
- \frac{1}{2} \left[ \hat{H}, \hat{\Lambda}(\underline{\pmb \lambda})  \right]_{+} &   = &  
        \left( - \Omega h({\pmb n }) \frac{ \partial }  {\partial \Omega}     
       - \sum_{x,y} {\pmb A} _{x,y} \frac{\partial}{\partial {\pmb n_{x,y}} } \right.     \\
 & + &\frac{1}{2} \sum_{\vec{i}, x,y,w,z} {\pmb B}^{(\vec{i})}_{x,y} {\pmb B}^{(\vec{i})}_{w,z}  
\frac{\partial^2}{\partial {\pmb n_{x,y}} \partial {\pmb n_{w,z}} }   \nonumber \\
 & +  &     \left.
\frac{1}{2} \sum_{\vec{i}, x,y,w,z} {\pmb C}^{(\vec{i})}_{x,y} {\pmb C}^{(\vec{i})}_{w,z}  
\frac{\partial^2}{\partial {\pmb n_{x,y}} \partial {\pmb n_{w,z}} }
 \right)  
  \hat{\Lambda}(\underline{\pmb \lambda}), \nonumber 
\end{eqnarray}
with
\begin{eqnarray}
h({\pmb n }) & = & {\rm Tr} \left(\hat{\Lambda} ({\pmb n}) \hat{H} \right), \label{ref_h}   \\
{\pmb A} & = & \frac{1}{2}  {\pmb n } \left( {\pmb T} -  U  {\pmb M} \right) 
 \overline{ {\pmb n } }  +
\frac{1}{2}  \overline {{\pmb n } } \left( {\pmb T} - U  {\pmb M} \right) 
{\pmb n }, \nonumber \\
{\pmb B}^{(\vec{i})}_{x,y} & = &  \sqrt{\frac{U}{2}} \sum_{\sigma,\sigma'}  
{\pmb n}_{x,(\vec{i},\sigma)}  {\pmb \sigma}_{\sigma,\sigma'}  
\overline{ {\pmb n} }_{(\vec{i},\sigma'),y},  
\nonumber \\
{\pmb C}^{(\vec{i})}_{x,y} & = &  \sqrt{\frac{U}{2}} \sum_{\sigma,\sigma'}  
\overline{ {\pmb n} }_{x,(\vec{i},\sigma)}  {\pmb \sigma}_{\sigma,\sigma'}  
 {\pmb n}_{(\vec{i},\sigma'),y}.  
\end{eqnarray}
In the above, $ \overline{ {\pmb n}   } = {\pmb 1}- {\pmb n}  $ and 
\begin{eqnarray}
\nonumber 
{\pmb M}_{(\vec{i},\sigma), (\vec{j},\sigma')}  & = &   \delta_{\vec{i},\vec{j}}  \sum_{\eta,\eta'} 
 {\pmb n}_{(\vec{i},\eta),(\vec{i},\eta') } {\pmb \sigma  }_{\sigma,\sigma'} 
        {\pmb \sigma  }_{\eta,\eta'}  \\ 
& + &  
( \pmb{1/2} -  {\pmb n})_{(\vec{i},\eta),(\vec{i},\eta') }  {\pmb \sigma  }_{\sigma,\eta} 
{\pmb \sigma  }_{\eta',\sigma'}. 
\nonumber
\end{eqnarray}
Partial integration, under the assumption that boundary terms vanish, yields the Fokker-Planck 
equation for the probability distribution $P(\underline{\pmb \lambda},\tau)$: 
\begin{eqnarray}
\label{Fokkerplanck}
        \frac{\partial }{ \partial \tau} P ( \underline{\pmb \lambda},\tau) & &  =   
\left[ \frac{\partial }{\partial \Omega}  \Omega h(\pmb n)  + 
\sum_{x,y} \frac {\partial} {\partial {\pmb n}_{x,y} } {\pmb A }_{x,y} \right.  \\ 
& + & \frac{1}{2} \sum_{\vec{i}, x,y,w,z}   
\frac{\partial^2}{\partial {\pmb n_{x,y}} \partial {\pmb n_{w,z}} } 
{\pmb B}^{(\vec{i})}_{x,y} {\pmb B}^{(\vec{i})}_{w,z}  \nonumber \\
 & +  &    
\left. \frac{1}{2} 
\sum_{\vec{i}, x,y,w,z} \frac{\partial^2}{\partial {\pmb n_{x,y}} \partial {\pmb n_{w,z}} }
{\pmb C}^{(\vec{i})}_{x,y} {\pmb C}^{(\vec{i})}_{w,z}   \right]
 P(\underline{\pmb \lambda}, \tau).  \nonumber
\end{eqnarray} 
The form of the diffusion matrices, 
$ {\cal \pmb D}^{C}_{(x,y), (w,z)} = 
\sum_{\vec{i}} {\pmb C}^{(\vec{i})}_{x,y} {\pmb C}^{(\vec{i})}_{w,z} $ 
and
$ {\cal \pmb D}^{B}_{(x,y), (w,z)} = \sum_{\vec{i}} {\pmb B}^{(\vec{i})}_{x,y} {\pmb B}^{(\vec{i})}_{w,z} $ 
is important. It  depends on the manner in which we have written the Hubbard interaction term, or in other
words on the choice of the diffusion gauge. The fact that the diffusion matrices factor out as above allows 
us to formulate the  SDE.   Furthermore, the fact that 
${\pmb B}^{\vec{i}} $  and $ { \pmb C}^{\vec{i}} $ are real  for real values of  ${\pmb n}$ will lead to 
positiveness of the weights.  The appropriate choice of diffusion gauges for  general Hamiltonians
is considered in Appendix \ref{diffusion_gauge}.

The assumption of vanishing boundary terms is essential to justify the approach and  
boils down to the requirement that the probability  $P( \underline{\pmb \lambda}, \tau)$ has 
tails decaying sufficiently fast as $|\vec{\lambda} | \rightarrow \infty$.
At this point, one has to recall that the  Gaussian basis  
is overcomplete, such that different probability distributions $P( \underline{\pmb \lambda}, \tau)$ 
will yield the the same density matrix.    This degree of freedom is reflected in  a stochastic 
gauge invariance which is reviewed in Appendix \ref{Stochastic_gauges}. Hence, even if
boundary terms appear the hope remains of eliminating them by  an appropriate stochastic gauge choice. 

To proceed, let us  assume that we can neglect the boundary terms.  The Fokker-Planck equation can conveniently be  transformed into 
an  Ito SDE \cite{Zinn-Justin}, 
\begin{eqnarray}
\label{SDE}
 {\rm d } {\Omega}  &  = & - \Omega h({\pmb n}) {\rm d} \tau, \\
 {\rm d } {\pmb n} & = & - {\pmb A } {\rm d} \tau +  
   \sum_{\vec{i}} {\pmb B}^{(\vec{i})} {\rm d} W_{\vec{i}} + 
   \sum_{\vec{i}} {\pmb C}^{(\vec{i})} {\rm d } W'_{\vec{i}} ,\nonumber
\end{eqnarray}
with Wiener increments $ \langle {\rm d} W_{\vec{i}} \rangle =\langle {\rm d} W'_{\vec{i}} \rangle   
=  \langle {\rm d} W_{\vec{i}}   {\rm d} W'_{\vec{j}} \rangle = 0$,   and 
$\langle {\rm d} W'_{\vec{i}}   {\rm d} W'_{\vec{j}} \rangle = 
\langle  {\rm d}W_{\vec{i}}   {\rm d} W_{\vec{j}} \rangle =
{\rm d} \tau \delta_{\vec{i},\vec{j} }$. 
Eq.~(\ref{SDE}) describes the time evolution of walkers in the space of Gaussian operators. 
At $\tau = 0$, $\rho(\tau=0) \propto \pmb 1$ such that all the Walkers can be parameterized by 
$\underline{{\pmb \lambda}} = ( 1, {\pmb { 1/2}} )$.
At imaginary time $\tau$ they are distributed according to 
$P(\underline{ {\pmb \lambda}},\tau)$ so that we have access to the density matrix.  
In particular, any equal time observable is given by: 
\begin{eqnarray}
        \langle \hat{ O} \rangle \simeq \frac 
   {\sum_i    {\rm Tr } \left[ \hat {\Lambda} 
      ( \underline{ \pmb \lambda}_{i}  ) \hat{O} \right] }
   {\sum_i {\rm Tr } \left[ \hat {\Lambda} ( \underline{ \pmb \lambda}_{i}  ) \right] }, 
\end{eqnarray}
where the sum runs over the set of walkers generated by the SDE.  Since Wick's theorem applies 
for a single Gaussian operator the numerator of the above equation may easily be calculated. 

As apparent from Eq. (\ref{SDE})  the weight of a Walker at imaginary time $\tau$ reads
\begin{equation} 
  \Omega(\tau) =   e^{ - \int_{0}^{\tau} {\rm d} \tau' h({\pmb n}(\tau') ) }.
\end{equation}
Since the ``equal time Green functions'', ${\pmb n}$,  are real, $h({\pmb n})$ is real and the weight remains 
positive. Hence the algorithm shows no explicit manifestation of the sign problem. 
However, the weights  grow exponentially with imaginary time thus 
yielding  an exponential increase in the  variance. To circumvent this problem, we have adopted 
the reconfiguration  scheme  proposed in Ref.~\cite{Calandra98}. In this approach the population  of 
walkers is kept constant. Walkers with large weights are cloned and those with small weights suppressed in 
such a way that in the large  population limit  the density matrix remains invariant.  Finally, after 
reconfiguration the weights of all walkers is equal to their average. 
\begin{figure}[t]
\begin{center}
\includegraphics[width=.4\textwidth,height=0.2\textheight]{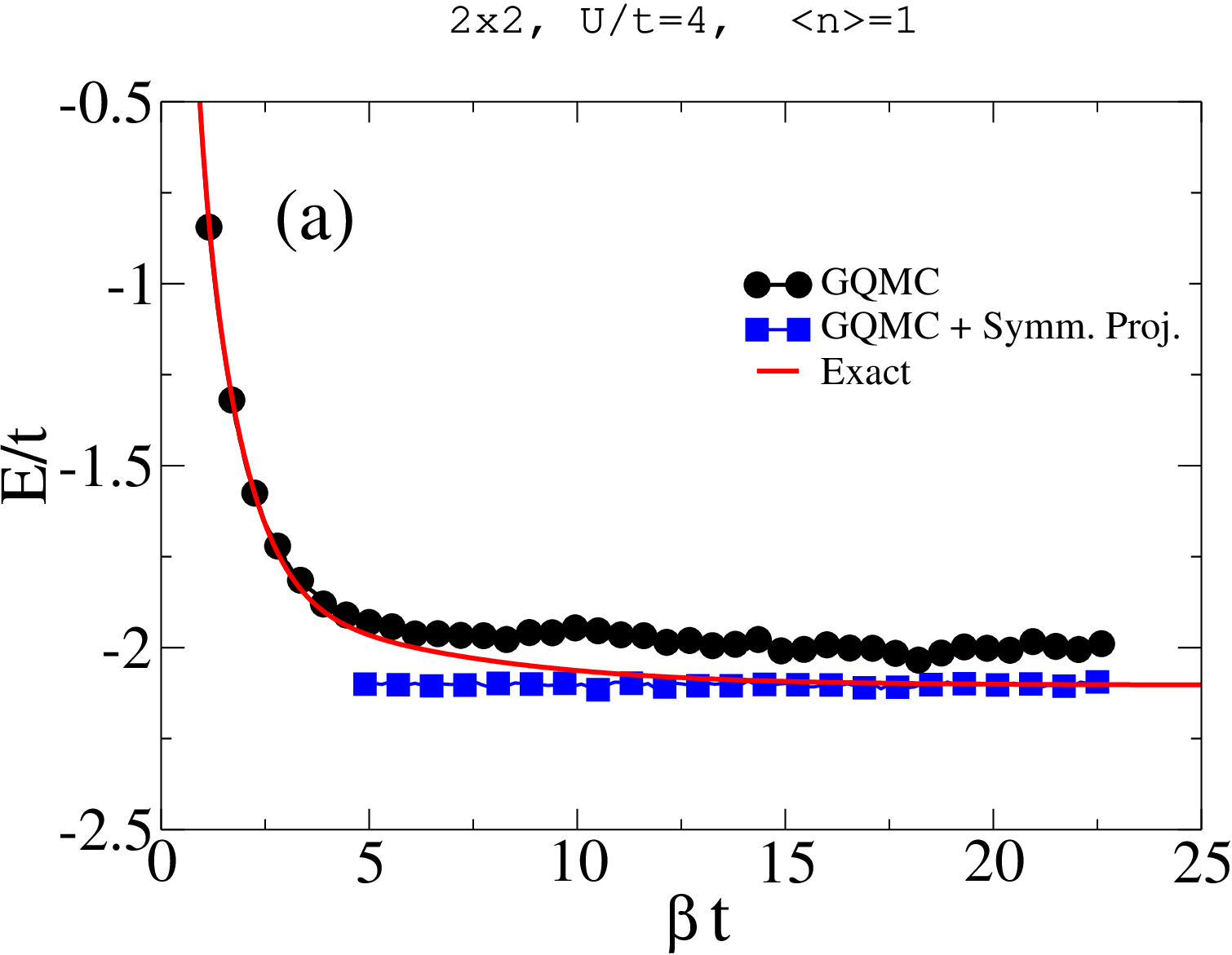} \\ 
\vspace{0.25cm}
\includegraphics[width=.4\textwidth,height=0.2\textheight]{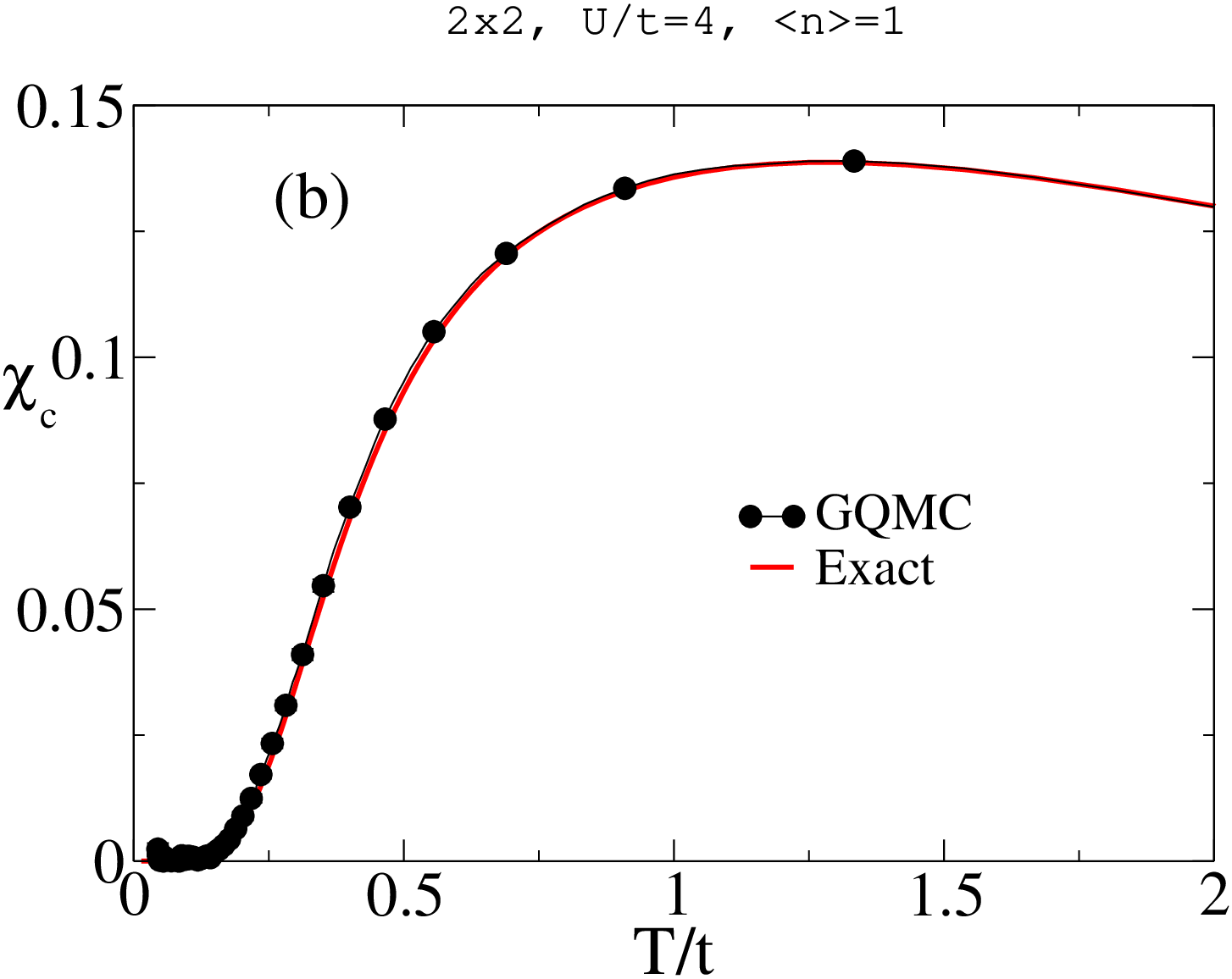} \\
\vspace{0.25cm}
\includegraphics[width=.4\textwidth,height=0.2\textheight]{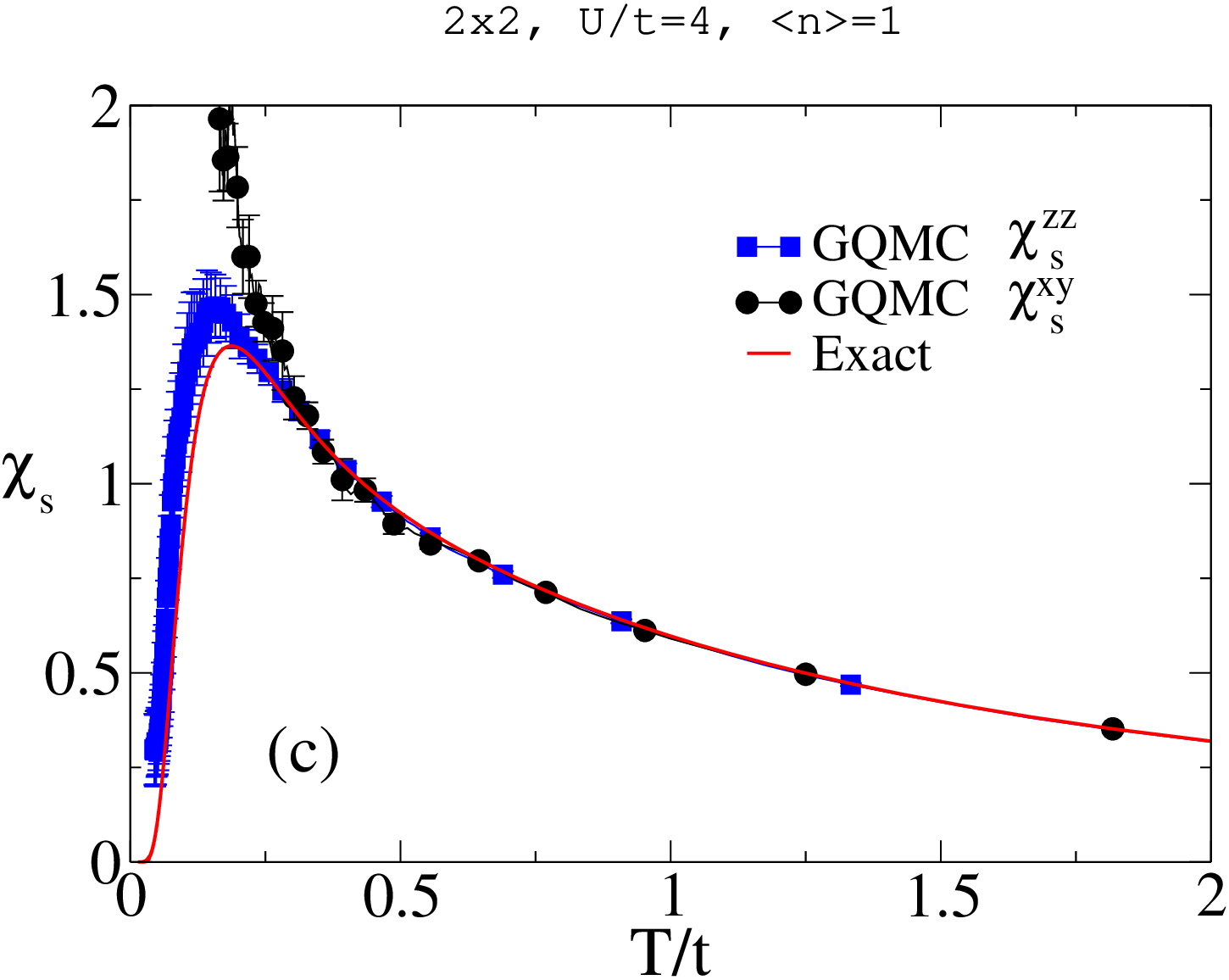} \\
\end{center}
\caption[]{ a) Energy as a function of inverse temperature as obtained from   exact diagonalization 
(solid line), from the GQMC (bullets) and from the GQMC supplemented by the
symmetry projection (squares).  Here we have projected onto the total spin $s=0$ state and 
$d$-wave lattice symmetry. 
(b) The charge susceptibility and (c) the longitudinal and transverse spin susceptibilities. 
In the above, we  use 60'000 walkers, an imaginary time step of $ \Delta \tau t = 0.0001  $, 
an explicit Euler scheme with adaptive time step,  and ${\pmb \sigma} = {\pmb \sigma}^z $ in Eq. 
(\ref{Hubbard})  }
\label{L2.fig}
\end{figure} 

We can now test the accuracy of the method on a $2 \times 2$ Hubbard model (See Fig.~\ref{L2.fig}). As 
apparent from Fig.~\ref{L2.fig}(a)  at {\it high} temperatures the  GQMC result   for the energy
(bullets in Fig. \ref{L2.fig}(a)) compares well with the exact result  (solid line). However at low 
temperatures there is a systematic deviation.  We have failed to 
account for this mismatch by i) enhancing the number of walkers  ii)  using
different schemes for the integration of the SDE iii)  varying the imaginary time step iv) 
setting ${\pmb \sigma}$ to $ {\pmb \sigma}_x $  instead of $ {\pmb \sigma}_z $ in Eq. (\ref{Hubbard})  and 
finally v) using different stochastic gauges (see Appendices
\ref{Stochastic_gauges} and \ref{diffusion_gauge}). Adding additional noise terms by means of diffusion gauges only seemed to make things worse. Concerning point ii) we tried implicit and explicit Euler schemes and a higher order Milstein integrator \cite{Kloeden}. As the latter
changes the order of the algorithm from $O(N^3)$ to $O(N^4)$ without reducing the systematic error, we prefer the Euler schemes. 

To acquire more insight into the origin of the mismatch, we compute the charge and spin 
susceptibilities: 
\begin{eqnarray}
        \chi_c  & = & \frac{\beta}{N} \left( \langle {\hat N }^2  \rangle -  
\langle {\hat N }  \rangle^2 \right)  \\
        \chi_s^z & = & \frac{\beta}{N} \left( \langle {\hat S }_z^2  \rangle -  
\langle {\hat S}_z  \rangle^2 \right)  \nonumber \\
        \chi_s^{xy} & = & \frac{1}{2} \frac{\beta}{N}  
 \left( \langle {\hat S }_x^2  \rangle -  \langle {\hat S}_x  \rangle^2 \right) + 
\frac{1}{2} \frac{\beta}{N}  
 \left( \langle {\hat S }_y^2  \rangle -  \langle {\hat S}_y  \rangle^2 \right)  \nonumber 
\end{eqnarray}
where
${\hat N }  = \sum_{\vec{i}} \hat{\pmb c}^{\dagger}_{\vec{i}} \hat{\pmb c}_{\vec{i}} $ and 
${\hat S }_\alpha =  \sum_{\vec{i}} \hat{\pmb c}^{\dagger}_{\vec{i}} {\pmb \sigma}^{\alpha} 
\hat{\pmb c}_{\vec{i}} $. Those quantities are plotted in Fig. \ref{L2.fig}(b),(c). As apparent, 
the charge susceptibility as well as $\chi_s^z$ follow rather precisely  the exact result.  On the 
other hand, $\chi_s^{xy}$ diverges as $1/T$ thereby signaling that the low temperature density 
matrix has non-vanishing  overlaps with  $s>0$ spin sectors.   To 
solve this problem so as to produce accurate ground state results we propose to implement symmetry 
projection schemes.

\section{Symmetry projections}
\label{Symm_proj.sec}

        Here we will assume that the low temperature density matrix has a large {\it overlap} with the 
ground state density matrix and  a {\it small } admixture of excited states.  
If this assumption is correct, then projection onto the
symmetry sector of the ground state will filter out the excited states and produce  
an accurate estimate of low temperature properties. Let us note that symmetry projection schemes 
have  been used successfully in the framework of the path-integral renormalization group approach 
where the ground state wave function is 
approximated by a sum  of Slater determinants \cite{Mizusaki04}. 
Here, we first review the mathematics of 
symmetry projections and then show how to implement them in the context of the GQMC. 

Let us first consider finite groups  with elements $R$ and irreducible representations 
${\cal D}^{\alpha}(R)$. Group theory then tells us that
\begin{equation}
\label{orthogonality}
        \sum_{R} {\cal D}^{\alpha}_{i,j}(R)^{\dagger} 
         {\cal D}^{\beta}_{i',j'}(R)  = \frac{\sum_{R}}{l_{\alpha}}
\delta_{\alpha,\beta} \delta_{i,i'} \delta_{j,j'},
\end{equation} 
where $l_{\alpha}$ corresponds to the dimension of the representation. 
For continuous groups, the sum has to be replaced by the  invariant integral: 
$\sum_{R} \rightarrow  \int {\rm d } R $ \cite{Wigner}.

To show how symmetry projections rely 
on the above identity let us first consider the group of translations by lattice vectors $\vec{R}$. 
\begin{eqnarray}
       & & {\hat T } (\vec{R}) \hat{c}^{\dagger}_{\vec{i},\sigma}       {\hat T}(\vec{R})^{-1} = 
         \hat{c}^{\dagger}_{\vec{i} + \vec{R},\sigma},
\end{eqnarray}
with  ${\hat T}(\vec{R}) = e^{ i \vec{R} \cdot \sum_{\vec{p},\sigma} \vec{p} \hat{c}^{\dagger}_{\vec{p},\sigma}
\hat{c}_{\vec{p},\sigma} }$ and 
$\hat{c}^{\dagger}_{\vec{p},\sigma} = \frac{1}{\sqrt{N}} \sum_{\vec{i}} e^{i \vec{p} \cdot \vec{i} }
\hat{c}^{\dagger}_{\vec{i},\sigma}$.
In the above, $N$ denotes the number of lattice sites.   Since the group of translations is an Abelian 
group, the irreducible representations are one-dimensional and labeled by the total momentum $\vec{K}$.
Classifying states in Fock space according to their total momentum, 
$\vec{K}$, yields  $ {\cal D}^{\vec{K}} (\vec{R}) = 
\langle \vec{K}, \alpha_{\vec{K}} | \hat{T}(\vec{R}) | \vec{K}, \alpha_{\vec{K}} \rangle  \equiv 
e^{i \vec{R} \cdot \vec{K} }$.  Here 
$ 1 = \sum_{\vec{K},\alpha_{\vec{K}}}| \vec{K}, \alpha_{\vec{K}} \rangle \langle \vec{K}, 
\alpha_{\vec{K}} |$, 
where $\alpha_{\vec{K}}$ labels all the states in  Fock space with total momentum $\vec{K}$.
The projection operator onto the Hilbert space with total momentum $\vec{K}_0$ reads:  
\begin{equation}
\hat{P}_{\vec{K_0}} = \frac{1}{N}  
       \sum_{\vec{R}} \langle \vec{K}_0 | \hat{T}(\vec{R}) | \vec{K}_0 \rangle^{\dagger} 
              \hat{T}(\vec{R}).
\end{equation}
This  expression may readily be verified:
\begin{eqnarray}
      & & P_{\vec{K_0}} | \Psi \rangle  =  \\
     & &  \sum_{\vec{K},\alpha_{\vec{K}}}  \overbrace{
        \frac{1}{N} 
       \sum_{\vec{R}} \langle \vec{K}_0 | \hat{T} (\vec{R}) | \vec{K}_0 \rangle^{\dagger} 
            \langle \vec{K}, \alpha_{\vec{K}} |  \hat{T}(\vec{R}) | \vec{K}, \alpha_{\vec{K}} \rangle}^{
            \delta_{ \vec{K}, \vec{K}_0 } } \nonumber \\
        & &  \times  \langle \vec{K}, \alpha_{\vec{K}}  | \Psi \rangle | \vec{K}, \alpha_{\vec{K}} \rangle  
        =  \sum_{\alpha_{\vec{K}_0}} \langle \vec{K}_0, \alpha_{\vec{K}_0}  | \Psi 
    \rangle | \vec{K}_0, \alpha_{\vec{K}_0} \rangle.  \nonumber
\end{eqnarray}

Within the very same framework,  we can define the projection on the Hilbert space with total spin $s$.  
We first parameterize the rotations in terms of  the  Euler angles, 
$ {\pmb \omega }= ( \alpha, \beta, \gamma )$,  such that  with 
\begin{equation}
        \hat{T} ( {\pmb \omega} ) = 
   e^{ i \alpha {\hat S}^z }
   e^{ i \beta  {\hat S}^y }
   e^{ i \gamma {\hat S}^z } 
\end{equation}
a spinor transforms as: 
\begin{equation}
    \hat{T} ( {\pmb \omega} ) {\pmb c}_{\vec{i}}^{\dagger}  \hat{T}^{-1} ( {\pmb \omega} ) 
=   {\pmb c}_{\vec{i}}^{\dagger}  e^{ i \frac{\alpha}{2} {\pmb \sigma}^z}
   e^{ i \frac{\beta}{2}  {\pmb \sigma}^y }
   e^{ i \frac{\gamma}{2} {\pmb \sigma}^z }. 
\end{equation}
Here, ${\hat S}^z$ corresponds to the total z-component of spin,  $ \sum_{\vec{i}} 
\frac{1}{2}  {\pmb c}^{\dagger}_{\vec{i}} {\pmb \sigma}^z {\pmb c}_{\vec{i}} $,   and a 
similar definition holds for ${\hat S}^y$.   Using Eq.~(\ref{orthogonality}) and noting that ${\cal D}^{s}_{m,m'}({\pmb \omega})  = 
\langle s,m | \hat{T} ( {\pmb \omega} ) | s,m' \rangle $  where the quantum numbers $m, m'$ denote
the z-component of spin, the projection onto the Hilbert space with definite spin $s$, and 
vanishing z-component of spin  reads: 
\begin{equation}
        {\hat P}_{s} =  \frac{2 s + 1}{ \int {\rm d} {\pmb \omega} }
 \int {\rm d} {\pmb \omega} \langle s,0 | \hat{T} ( {\pmb \omega} ) | s,0 \rangle^{\dagger}
 \hat{T} ( {\pmb \omega} ).
\end{equation}
Since we have chosen to parameterize rotations in terms of Euler angles the invariant 
integral reads: 
$ \int {\rm d} {\pmb \omega}  = \int_{0}^{2\pi} {\rm d} \alpha \int_{0}^{\pi} {\rm d} 
\beta \sin(\beta)  \int_{0}^{2 \pi} {\rm d} \gamma $ and
\begin{equation}
        \langle s,0 | \hat{T} ( {\pmb \omega} ) | s,0 \rangle = P_{s}(\cos(\beta)),
\end{equation}
where  $P_{s}$ denotes the $s^{th}$ Legendre polynomial. 

Since the GQMC method is a grand canonical approach we have equally implemented 
projection onto  fixed particle number.  To this purpose, we define the  gauge transformation:  
\begin{equation}
        \hat{T}(\phi) = e^{i \phi \sum_{i}  {\pmb c}^{\dagger}_{\vec{i}} {\pmb c}_{\vec{i} } }  
\end{equation}
such that $ \hat{T}(\phi) {\pmb c}_{\vec{i}}^{\dagger} \hat{T}^{-1}(\phi) =
e^{i\phi} {\pmb c}_{\vec{i}}^{\dagger} $.  Projection onto a given particle number  sector  then reads: 
\begin{equation}
        \hat{P}_N  = \frac{1}{2 \pi} \int_{0}^{2 \pi} \langle N | \hat{T}(\phi) | N \rangle^{\dagger} 
\hat{T}(\phi). 
\end{equation} 

Finally, we have implemented the $C_4$ lattice symmetries to classify states according to 
i)  s-wave: even under parity and  $\pi/2$ rotations 
ii) d-wave: even under parity and odd under  $\pi/2$ rotations  and 
iii) $p_x + ip_y$: odd under parity and  acquires a phase factor $ e^{i \pi/2}$ under $\pi/2$ 
rotations.   We denote this projection by $ \hat{P}_\text{latt} $. 

Since the Hubbard Hamiltonian is invariant under lattice vector translations, spin rotations, 
gauge transformations, $\pi/2$ rotations, the ground state density matrix will have definite 
momentum, spin, particle number and lattice symmetry.  Our aim is now to project the 
density matrix produced by the GQMC onto
a given symmetry sector and then use the projected density matrix,
\begin{equation} 
\hat{P} \hat{\rho} \hat{P}^{\dagger},  
\end{equation}
to compute observables.
Here, $ \hat{P} $ is a product of all or only some of the 
above symmetry projectors with general form, 
\begin{equation}
                \hat{P} = \int {\rm d} {\pmb x} g({\pmb x})  \hat{T}({\pmb x}),
\end{equation} 
where $\hat{T}$ is unitary and $ \hat{P}^{\dagger}  =  \hat{P} $. 

\begin{table}[h]
\begin{tabular}{|c|c|c|}
\hline
 $2\times 2, U/t=4 $       & GQMC +  Symm. Proj. & Exact   \\
 $ \langle n \rangle =1$   & $s=0$, $d$-wave &   \; \\
\hline
   Energy/t     &   $-2.1021   \pm   0.0007 $  &  $ -2.1026 $ \\
  $ S(\pi,\pi)$ &   $ 2.1933   \pm   0.0010 $  &  $  2.1947 $ \\
  $ N(\pi,\pi)$ &   $ 0.2667   \pm   0.0004 $  &  $  0.2664 $  \\
\hline
\end{tabular}
\caption{ GQMC with symmetry projection for the $2 \times 2$ half-filled Hubbard model at 
$U/t=4$. Here we have projected onto the d-wave  and  spin-singlet Hilbert spaces. 
To impose the   spin projection we have to integrate over the three Euler angles. This integration 
is done numerically by replacing the three dimensional integral by a Riemann sum over $5^3$ points. 
The thus produced systematic error is not included in the error bars.  The results and error bars stem 
from averaging  the data over imaginary time (squares in Fig. \ref{L2.fig}). }
\label{Table_L2.dat}
\end{table}

To simplify the calculation, we will assume that the observable  $ \hat{O}$ commutes with $\hat{P}$: 
\begin{equation} 
\left[ \hat{P}, \hat{O} \right]_{-}   = 0,
\end{equation}
such that
\begin{eqnarray}
\label{Symm_proj}
        \langle \hat{O} \rangle_P = \frac{ {\rm Tr} \left[ \hat{P} \hat{\rho} \hat{P} O \right]  } 
                                         { {\rm Tr} \left[ \hat{P} \hat{\rho} \hat{P}  \right]   } 
                           = \frac{ {\rm Tr} \left[ \hat{P} \hat{\rho}  \hat{O} \right]  } 
                                    { {\rm Tr} \left[\hat{P} \hat{\rho}  \right]   },
\end{eqnarray}
since $ \hat{P}^2 = \hat{P}$. Estimating the right hand side of the above equation boils down to the 
calculation of $\hat{P} \hat{\rho}$ where  
$\hat{\rho}  \approx \sum_{\underline{\pmb \lambda}}  
\hat{\Lambda}(\underline{\pmb \lambda} )$ and the sum runs over the walkers produced  by 
integrating the SDE.
Hence, using the result of  Appendix \ref{Unitary_transform}, 
\label{}
\begin{eqnarray}
\label{Symm_proj_1}
\langle \hat{O} \rangle_P & = &\frac{ \sum_{\underline{\pmb \lambda}} \int {\rm d} {\pmb x} g({\pmb x}) 
            {\rm Tr} \left[ \hat{T}({\pmb x}) \hat{\Lambda}(\underline{\pmb \lambda}) \hat{O} \right]  } 
            { \sum_{\underline{\pmb \lambda}} \int {\rm d} {\pmb x} g({\pmb x}) {\rm Tr} 
             \left[ \hat{T}({\pmb x}) \hat{\Lambda}(\underline{\pmb \lambda})  \right]   } \nonumber \\
        & = & \frac{ \sum_{\underline{\pmb \lambda}} \int {\rm d} {\pmb x} g({\pmb x}) 
            {\rm Tr} \left[  \hat{\Lambda}(\underline{\pmb \lambda}({\pmb x})) \hat{O} \right]  } 
            { \sum_{\underline{\pmb \lambda}} \int {\rm d} {\pmb x} g({\pmb x}) \Omega(\pmb{x}) },
\end{eqnarray}
where $ \hat{T}({\pmb x}) \hat{\Lambda}(\underline{\pmb \lambda}) = 
\hat{\Lambda}(\underline{\pmb \lambda}({\pmb x})) $. 

We test the above procedure on the $2 \times 2 $ lattice of Fig.~\ref{L2.fig}. As seen in  
Fig. \ref{L2.fig}(a) (solid squares) by projecting onto the spin-singlet and d-wave state we 
obtain a very accurate estimate of the ground state energy  already at $ \beta t = 5$. 
Averaging over subsequent  imaginary times  
yields the results  presented in Table \ref{Table_L2.dat}. It is important to note that not only 
the ground state energy is very well reproduced but that also reliable estimates for the spin and 
charge structure factors,
\begin{eqnarray}
S(\vec{q}) & = & \frac{4}{3N} \sum_{\vec{i},\vec{j}} e^{\vec{q} \cdot \left( \vec{i}-\vec{j} \right)} 
     \langle \hat{\pmb S}_{\vec{i}} \cdot \hat{\pmb S}_{\vec{j}} \rangle, \\
N(\vec{q}) & = & \frac{1}{N} \sum_{\vec{i},\vec{j}} e^{\vec{q} \cdot \left( \vec{i}-\vec{j} \right)} 
     \langle \hat{n}_{\vec{i}} \cdot \hat{n}_{\vec{j}} \rangle,  \nonumber
\end{eqnarray}
are obtained. 

\section{Accuracy tests}  
\label{Accuracy_tests}

\begin{table}[h]
\begin{center}
                $  U/t=4, t'/t = 0,  \langle n \rangle  = 1 $
\end{center}
\begin{tabular}{|c|c|c|}
\hline
 $ L = 4  $   & GQMC +  Symm. Proj. & Exact   \\
 $ \;     $   & $s=0$,  $\vec{P}=0$   & \;       \\
\hline
   Energy/t     &   $-13.630  \pm    0.016  $  &  $ -13.6224 $ \\
  $ S(\pi,\pi)$ &   $  3.66   \pm    0.013  $  &  $  3.64  $ \\
  $ N(\pi,\pi)$ &   $  0.386  \pm    0.001  $  &  $  0.385 $  \\
\hline 
\hline
 $ L = 6  $   & GQMC +  Symm. Proj. &  PQMC   \\
 $ \;     $   &  $s=0$, $\vec{P}=0$   &    \; \\ 
\hline
   Energy/t     &   $ -30.87  \pm    0.04   $  &  $ -30.87 \pm 0.02 $ \\
  $ S(\pi,\pi)$ &   $  5.86   \pm    0.05   $  &  $  5.82  \pm 0.03 $ \\
  $ N(\pi,\pi)$ &   $  0.400  \pm    0.004  $  &  $  0.418 \pm 0.025 $  \\
\hline 
\end{tabular}

\caption{Comparison between GQMC and benchmark results
for the $4 \times 4 $ and $6 \times 6$  Hubbard model.
For both parameter sets, we project 
onto  total spin  $s=0$  and total momentum $\vec{P} = 0$. 
The $L=4$ ($L=6$) simulations were carried out with $12'000$ ($6'000$) walkers, 
an explicit Euler scheme and an  imaginary time step $ \Delta \tau t = 0.0005$  ($\Delta \tau t = 0.001$).
The exact diagonalization results for the $L=4$ lattice stem from Ref. \cite{Parola91}. 
For the $L=6$ lattice we compare with the  auxiliary field projector QMC (PQMC) algorithm.  } 
\label{Table_ns.dat}
\end{table}

\begin{figure}[h]
\begin{center}
\includegraphics[width=.4\textwidth,height=0.23\textheight]{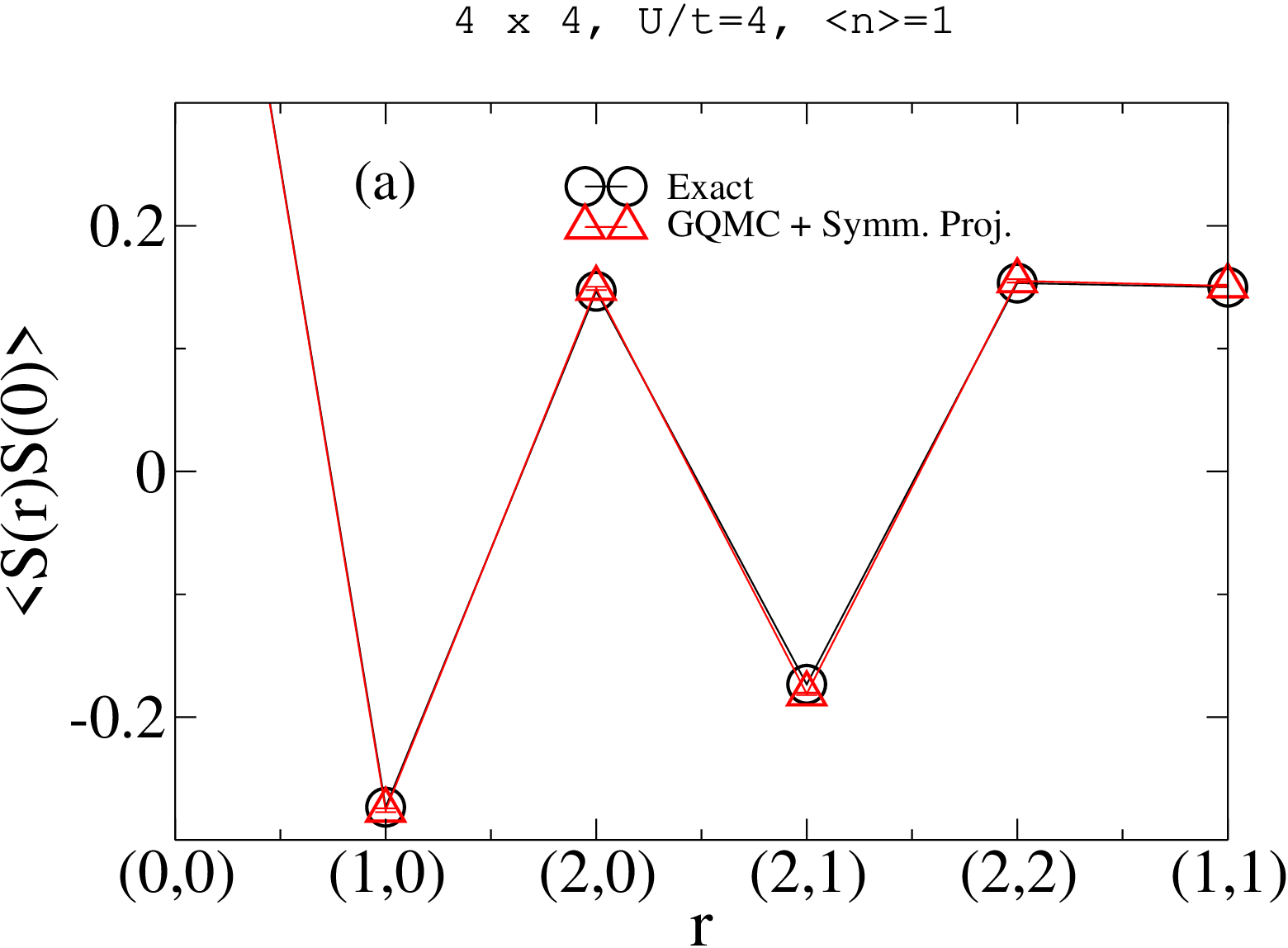} \\ 
\vspace{0.25cm}
\includegraphics[width=.4\textwidth,height=0.23\textheight]{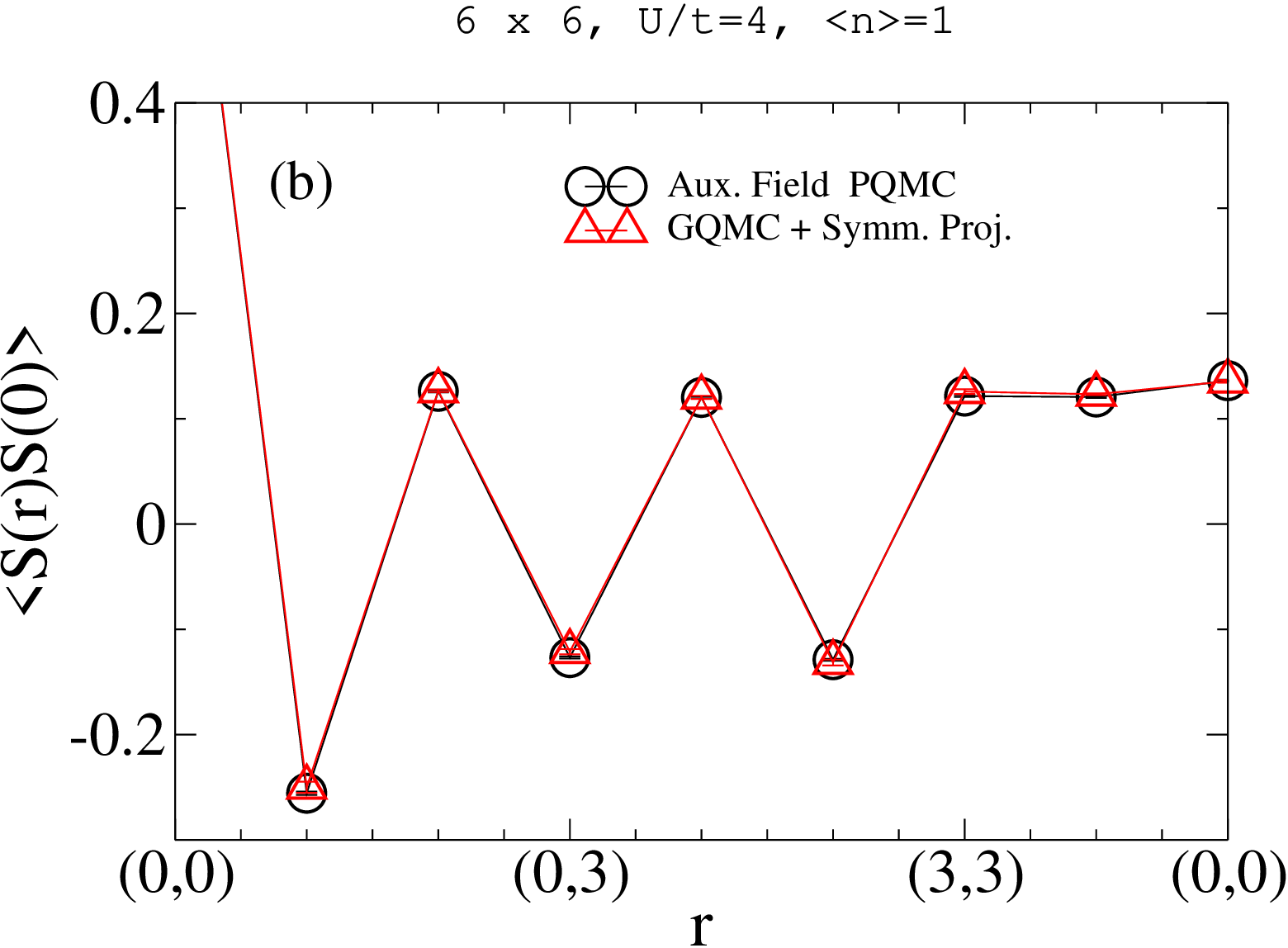} 
\end{center}
\caption[]{ Real space spin-spin correlations as obtained from the GQMC and comparison with 
benchmark results.  See caption of Table \ref{Table_ns.dat} for details of the 
GQMC simulations. }
\label{Spin_ns.fig}
\end{figure} 

Here we provide further tests triggered at assessing the accuracy of the  method. 
We first consider systems where the sign  problem is absent in auxiliary field QMC methods, that 
is, the particle-hole symmetric  Hubbard model.  Table \ref{Table_ns.dat} presents results  at half-filling
for both $4 \times 4 $ and $6 \times  6 $ lattices.  In both cases, one sees that the agreement with 
benchmark results (exact diagonalization for the $4\times 4 $ lattice  and auxiliary field projector 
QMC (PQMC) for the $6 \times 6$ lattice) is excellent.  Furthermore, the real space spin-spin 
correlations agree
very well with the benchmark results (see Fig. \ref{Spin_ns.fig}).

\begin{table}[h]
\begin{tabular}{|c|c|c|}
\hline
 $  U/t=4, t'/t= 0          $   &  GQMC +  Symm. Proj. &      Exact   \\
 $ \langle n \rangle =0.625 $   &  $s=0$, $s$-wave, $N = 10$  &  \; \\
\hline
   Energy/t     &   $ -19.576     \pm    0.012585  $  &  $ -19.584   $ \\
  $ S(\pi,\pi)$ &   $  0.737      \pm    0.002  $     &  $  0.73     $ \\
  $ N(\pi,\pi)$ &   $  0.5075     \pm    0.001  $     &  $  ----     $  \\
\hline
\hline
 $  U/t=8, t'/t=-0.3    $   &  GQMC +  Symm. Proj.           &    Exact   \\
 $ \langle n \rangle =1 $   &  $s=0$, $\vec{P} = 0$, $s$-wave   &   \;  \\
\hline
   Energy/t     &   $ -8.498   \pm   0.012  $  &  $ -8.4884  $ \\
  $ S(\pi,\pi)$ &   $ 5.09     \pm   0.07   $  &  $ 4.985 $ \\
  $ N(\pi,\pi)$ &   $ 0.191   \pm   0.004   $  &   $  0.1920 $  \\
\hline
\hline
 $  U/t=8, t'/t=-0.3    $       & GQMC +  Symm. Proj. &     Exact   \\
 $ \langle n \rangle =0.875 $   & $s=0$, $\vec{P} = 0$, $N = 14$ &  \;  \\
\hline
   Energy/t     &   $ -12.01  \pm   0.40  $  &   $  -12.50293   $ \\
  $ S(\pi,\pi)$ &   $  0.941  \pm   0.17  $  &   $  0.964776 $ \\
  $ N(\pi,\pi)$ &   $  0.266  \pm   0.01  $  &   $  0.27962 $  \\
\hline

\end{tabular}
\caption{ Comparison between GQMC and exact diagonalization results. Here we have used $12'000$ walkers 
and a time step of $\Delta \tau t = 0.0005$.  The  GQMC is a grand-canonical simulation. Hence 
in cases where charge fluctuations are not negligible  we 
project onto fixed particle number Hilbert spaces so as to allow comparison with exact diagonalization
results. }
\label{Table_s.dat}
\end{table}

\begin{figure}[h]
\begin{center}
\includegraphics[width=.4\textwidth,height=0.23\textheight]{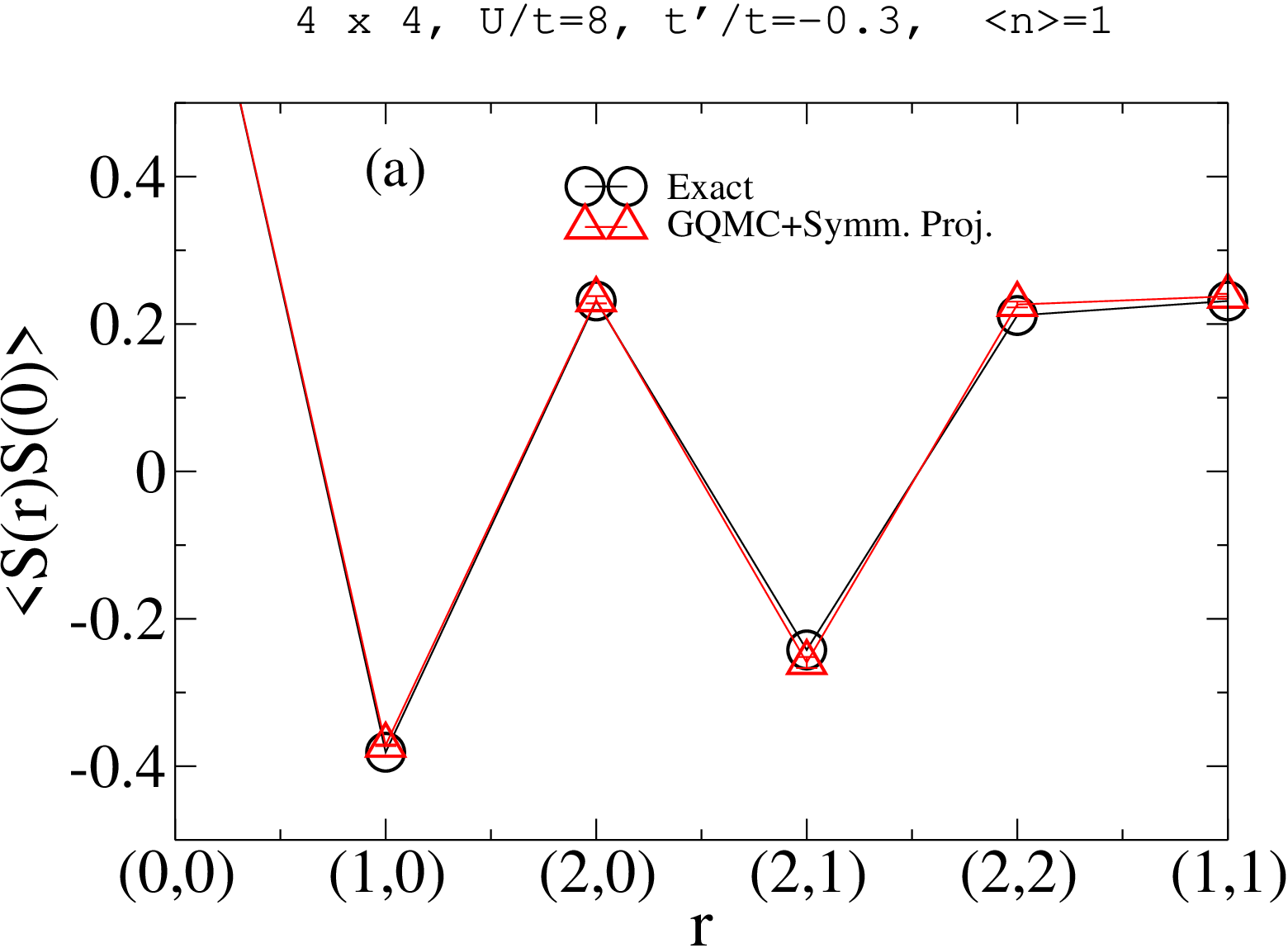} 
\vspace*{0.1cm}
\includegraphics[width=.4\textwidth,height=0.23\textheight]{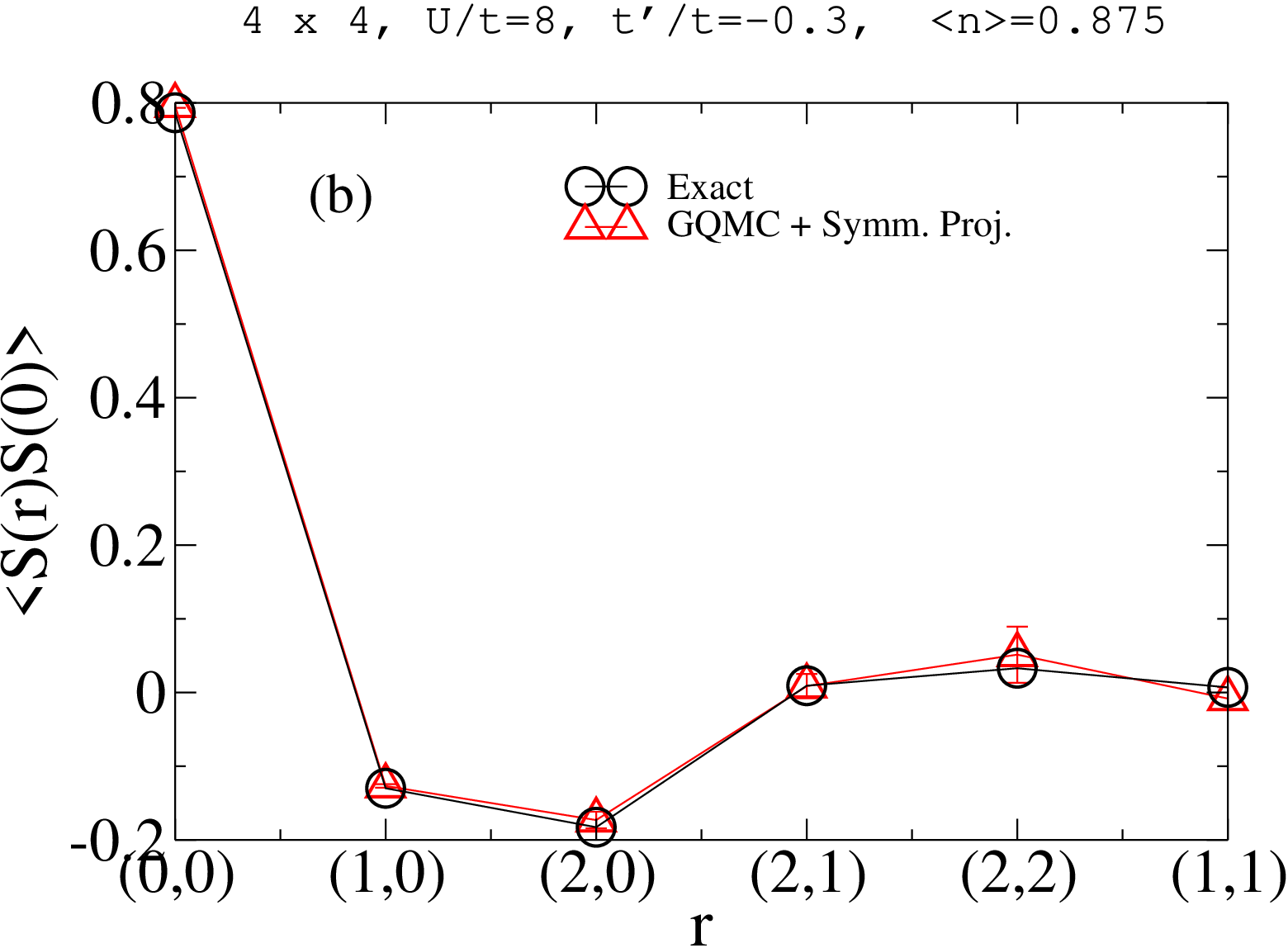} 
\end{center}
\caption[]{ Real space spin-spin correlations as obtained from the GQMC and comparison with 
exact-diagonalization results.  See caption of Table \ref{Table_s.dat} for details of the 
GQMC simulations. }
\label{Spin_s.fig}
\end{figure}

The crucial point is to show that in situations where the sign problem plagues the  auxiliary field
QMC, the Gaussian approach remains accurate. Table  \ref{Table_s.dat}  presents three data sets 
where the sign problem in the auxiliary field approach varies from mild to very severe. \\
i) Let us start with 
the $4 \times 4 $ Hubbard model with nearest neighbor hopping,  $t$,  and  
$ \langle n \rangle = 10/16$. The agreement between 
the GQMC and exact diagonalization  is excellent. It is worth pointing out that in this 
specific case,  the  GQMC results with and without symmetry projections are  identical meaning that 
the GQMC automatically produces the ground state density matrix with correct symmetries.  We believe that 
this is due to the fact that at this large doping away from half-filling the ground state  is 
very well described by a paramagnetic mean field solution. Such a mean field solution is exactly 
reproduced by the GQMC approach.   \\
ii) Our second example is the  half filled $4 \times 4 $ frustrated Hubbard model at $U/t=8$. Here 
frustration stems from a next-nearest neighbor hopping $t'/t  = -0.3$. 
Both Table~\ref{Table_s.dat} and Fig.~\ref{Spin_s.fig}(a)  show that we obtain excellent agreement 
with the exact result.  
Note that for those model parameters the finite temperature auxiliary field approach 
has an  average sign of $\langle {\rm sign} \rangle \approx 0.2 $ at  $\beta t = 10$ and 
of $\langle {\rm sign} \rangle \approx 0.1 $ at $\beta t = 15$.  \\ 
iii) We now consider a parameter set which is out of reach for the auxiliary field approach, 
$U/t =8 $, $\langle n \rangle = 0.875 $ and $t'/t = -0.3$. Table \ref{Table_s.dat} shows that 
we are  capable of reproducing the exact results. However the fluctuations  and hence the error bars 
in the MC data are large  in comparison to the half-filled case.
Those  large fluctuations stem from the symmetry projection.  In particular,  the denominator in 
Eq. (\ref{Symm_proj}),  $ {\rm Tr} \left[ \hat{P} \hat{\rho} \right] $,  
is {\it small}  and has large  relative fluctuations.  In other words, the low temperature  density matrix 
(here we have propagated the walkers up to $\beta t = 40$)
produced by the GQMC still includes many excited states, and it is hard to filter out  ground 
state properties by imposing symmetries \footnote{ It is known that weakly doped Mott insulators have many 
competing  states or in other words  a dense spectrum  of low lying states.  Hence at finite but low 
temperatures, the density matrix  will contain excited states.}. 
Nevertheless, and as 
seen in Fig. \ref{Spin_s.fig}(b),  comparisons with exact diagonalization results show that 
we are capable of accurately reproducing the details of real space spin-spin correlation function. 

\section{Conclusions}
We have shown that the GQMC method produces inaccurate ground state properties since the numerical 
solution of the SDE fails to produce a low temperature density matrix with the symmetry properties 
of the Hamiltonian. To repair this sampling problem,  we propose to a posteriori  project the density 
matrix onto the symmetry sector of the  ground state.  We have shown ample non-trivial tests, 
including situations where  auxiliary field methods fail due to the sign problem,  
where this approach yields accurate and reliable results. Those results confirm the 
point of view that the low temperature density  matrix produced by  the GQMC  has a good  overlap with 
the exact zero temperature density matrix,  but that the GQMC density matrix contains excited states, because the symmetries are not correctly reproduced.  Those excited states are filtered out 
by the projection. 

There are many open questions which  deserve further work. In particular,  is it possible to  improve the 
sampling by incorporating aspects of the symmetry projections directly into the SDE?  Also, we have not 
yet fully  exploited the flexibility of the stochastic gauges.   It is at present not clear if, 
with a suitable choice of stochastic gauge, the here mentioned  symmetry problems may be solved. 

\acknowledgements
The calculations presented here were carried out on the IBM p690 cluster of
the NIC  in J\"ulich. We would like to thank this institution for allocation of CPU time.  We have 
greatly profited from  discussions with G.G. Batrouni, C. Br\"unger,  J. Corney, P. Drummond, W.P. Petersen
and D. Talay.
Many thanks to S. Capponi who provided  part of the exact diagonalization benchmark results. 
We acknowledge support by the Swiss National Science Foundation  and DFG.

\appendix
\section{Unitary transformation of a Gaussian operator}
\label{Unitary_transform}
In this appendix we show that:
\begin{eqnarray}
\label{Appendix_res}
  & &   e^{ i \hat{\pmb c}^{\dagger}{\pmb h } \hat{\pmb c} } \hat{\Lambda}(\underline{\pmb \lambda} ) 
 = \hat{\Lambda}(\tilde{\underline{\pmb \lambda}} ),
 \end{eqnarray}
with
\begin{eqnarray}
 (\tilde{ \pmb n}^{T} -1 )^{-1} &=& 
         \left[ ( e^{i {\pmb h}} -1 ) { \pmb n}^{T}  + {\pmb 1}   \right] ({ \pmb n}^{T} -1)^{-1},
\nonumber \\ 
 \tilde{\Omega} &=& \Omega \det \left[ ( e^{i {\pmb h}} -1 ) { \pmb n}^{T}  + {\pmb 1}   \right].\nonumber
\end{eqnarray}
Here, ${\pmb h}^{\dagger} = {\pmb h}$, $\tilde{\underline{\pmb \lambda}} = (\tilde{\Omega}, \tilde{\pmb n})$, 
and ${\underline{\pmb \lambda}} = ({\Omega}, {\pmb n})$.

Before showing the above, let us fist recall some  identities of the  Grassmann  algebra \cite{Negele}:
\begin{eqnarray}
\langle { \pmb  \xi} | { \pmb  \xi}' \rangle &=&  e^{\sum_x \xi_x^{\dagger} \xi'_x } 
\equiv e^{ {\pmb  \xi}^{\dagger} {\pmb \xi}' },  \nonumber \\
\langle { \pmb  \xi} |:A({\pmb c}^\dagger, {\pmb c}): | { \pmb  \xi}' \rangle   &=& 
A({\pmb \xi}^{\dagger},{\pmb \xi}') e^{ {\pmb \xi}^{\dagger} {\pmb \xi}'},   \nonumber \\
 {\pmb 1}  &=&  \int \underbrace{\prod_x {\rm d} \xi_x^{\dagger} {\rm d} \xi_x}_{
    \equiv {\cal D} {\pmb  \xi} }
 e^{-  {\pmb \xi}^{\dagger} {\pmb \xi} }  
|{ \pmb  \xi} \rangle \langle  { \pmb  \xi} |. 
\end{eqnarray}
Here $   \xi_x $ are Grassmann variables and $ |{ \pmb  \xi} \rangle $ fermion coherent states.  

In a first step it is convenient to transform $e^{ i \hat{\pmb c}^{\dagger} h \hat{\pmb c} } $ into a
normal ordered form. Since $h$ is hermitian,  
$ {\pmb  h}  = {\pmb U} {\pmb D} {\pmb  U}^{\dagger}$ with $ {\pmb D} $ a diagonal 
and ${\pmb U}$ unitary.   With the canonical transformation  
$ \hat{\pmb \gamma}^{\dagger}  =   \hat{\pmb c}^{\dagger} {\pmb  U }  $ we obtain: 
\begin{eqnarray}
        e^{ i \hat{\pmb c}^{\dagger} {\pmb h} \hat{\pmb c} } & = &  
        \prod_x e^{ i \hat{\gamma}^{\dagger}_x \hat{\gamma}_x D_x       } = 
        \prod_x \left[ 1 + \left(e^{i D_x} -1 \right)\hat{\gamma}^{\dagger}_x \hat{\gamma}_x \right]  
     \nonumber \\
   & = & \prod_x : e^{\left(e^{i D_x} -1 \right)\hat{\gamma}^{\dagger}_x \hat{\gamma}_x } :  
    = : e^{\sum_x \hat{\gamma}^{\dagger}_x \left( e^{iD_x} -1 \right)  \hat{\gamma}_x } : \nonumber \\
    & = & : e^{ \hat{\pmb c}^{\dagger} \left( e^{i {\pmb h} } -{\pmb 1} \right) \hat{\pmb c}} : 
\end{eqnarray} 

We can now compute the quantity $ e^{ i \hat{\pmb c}^{\dagger} {\pmb h} \hat{\pmb c} }  
: e^{ \hat{\pmb c}^{\dagger}  {\pmb B} \hat{\pmb c} }: $ where $ {\pmb B} $ is an arbitrary matrix:
\begin{eqnarray}
& & e^{ i \hat{\pmb c}^{\dagger} {\pmb h} \hat{\pmb c} }  
: e^{ \hat{\pmb c}^{\dagger}  {\pmb B} \hat{\pmb c} }:  
= : e^{ \hat{\pmb c}^{\dagger} \left( e^{i {\pmb h} } -1 \right) \hat{\pmb c}} : 
: e^{ \hat{\pmb c}^{\dagger}  {\pmb B} \hat{\pmb c} }: =  \nonumber \\
& & \int {\cal D} {\pmb \xi} {\cal D} {\pmb \eta}{\cal D} {\pmb \gamma} 
e^{- {\pmb \xi}^{\dagger} {\pmb \xi}  -  {\pmb \eta}^{\dagger}  {\pmb \eta}  - 
 {\pmb \gamma}^{\dagger}  {\pmb \gamma} } | { \pmb \xi} \rangle \langle { \pmb \xi} |: e^{ \hat{\pmb c}^{\dagger} \left( e^{i {\pmb h}} -1 \right) \hat{\pmb c}} :  \times \nonumber \\
& & \; \; \;  | { \pmb \eta} \rangle \langle { \pmb \eta} |  
: e^{ \hat{\pmb c}^{\dagger}  {\pmb B} \hat{\pmb c} }: 
| { \pmb \gamma} \rangle \langle { \pmb \gamma} | = \nonumber  \\
& & \int  {\cal D} {\pmb \xi} {\cal D} {\pmb \eta}{\cal D} {\pmb \gamma} 
e^{- {\pmb \xi}^{\dagger} {\pmb \xi}  -  {\pmb \eta}^{\dagger}  {\pmb \eta}  - 
 {\pmb \gamma}^{\dagger}  {\pmb \gamma} } | { \pmb \xi} \rangle  
e^{ {\pmb \xi}^{\dagger}  e^{i {\pmb h}}  {\pmb \eta}}  
e^{ {\pmb \eta}^{\dagger}  ( {\pmb B} + 1) {\pmb \gamma } } \langle { \pmb \gamma} |  = \nonumber \\
& & \int  {\cal D} {\pmb \xi} {\cal D} \tilde{\pmb \eta}{\cal D} {\pmb \gamma} 
e^{- {\pmb \xi}^{\dagger} {\pmb \xi}  -  \tilde{\pmb \eta}^{\dagger}  \tilde{\pmb \eta}  - 
 {\pmb \gamma}^{\dagger}  {\pmb \gamma} } | { \pmb \xi} \rangle  
e^{ {\pmb \xi}^{\dagger}    \tilde{\pmb \eta}} 
 e^{ \tilde{\pmb \eta}^{\dagger}  e^{i {\pmb h}} ( {\pmb B} + 1) {\pmb \gamma } } \langle { \pmb \gamma} |  = \nonumber \\
& & \int  {\cal D} {\pmb \xi} {\cal D} \tilde{\pmb \eta}{\cal D} {\pmb \gamma} 
e^{- {\pmb \xi}^{\dagger} {\pmb \xi}  -  \tilde{\pmb \eta}^{\dagger}  \tilde{\pmb \eta}  - 
 {\pmb \gamma}^{\dagger}  {\pmb \gamma} } | { \pmb \xi} \rangle   
  \langle  {\pmb \xi} | \tilde{\pmb \eta}  \rangle \times  \nonumber \\ 
 & &    \langle \tilde{\pmb \eta} |  :e^{ {\pmb c}^{\dagger}  
[e^{i {\pmb h}} ( {\pmb B} + 1) -1 ] {\pmb c} } : | {\pmb \gamma } \rangle \langle { \pmb \gamma} |  = 
 :e^{ {\pmb c}^{\dagger}  [e^{i {\pmb h}} ( {\pmb B} + 1) -1 ] {\pmb c} } : 
\end{eqnarray}
Here, we have carried out the substitution 
$ \tilde{\pmb \eta}  = e^{i {\pmb h}} {\pmb \eta} $, baring in mind
that $ e^{i{\pmb  h}}$ is unitary matrix. 

The  result of Eq. \ref{Appendix_res} follows from: 
\begin{eqnarray}
   & & e^{i \hat{\pmb c}^{\dagger} {\pmb h} \hat{\pmb c} } \hat{\Lambda(\underline{\pmb \lambda})} = 
        \Omega \det(1-{\pmb n}) e^{i \hat{\pmb c}^{\dagger} {\pmb h} \hat{\pmb c} } : 
    e^{-  \hat{\pmb c}^{\dagger}\left[ {\pmb 2}  + ( {\pmb n}^T - {\pmb 1} )^{-1}\right] \hat{\pmb c} }:  =
\nonumber \\
  & &   \Omega \det(1-{\pmb n}) 
  : e^{ -\hat{\pmb c}^{\dagger}\left[ {\pmb 1}  + e^{i {\pmb h}}( 1+ ({\pmb n}^T - {\pmb 1} )^{-1} )\right] \hat{\pmb c} }:  \equiv \nonumber \\
& &  \underbrace{\Omega \frac{\det(1-{\pmb n}) }{\det(1-\tilde{\pmb n}) } }_{\tilde{\Omega}} \;
   \underbrace{  {\det(1-\tilde{\pmb n}) }  : e^{-  \hat{\pmb c}^{\dagger}\left[ {\pmb 2}  + ( \tilde{\pmb n}^T - {\pmb 1} )^{-1}\right] \hat{\pmb c} }:  }_{ \hat{\Lambda}(\tilde{\pmb n}) }
\end{eqnarray}

\section{Drift gauges} 
\label{Stochastic_gauges}
Since the Gaussian operator basis is overcomplete, there are many probability distributions
$P(\underline{ \pmb \lambda}, \tau) $ which will result in the same density matrix. This degree 
of freedom on $P(\underline{ \pmb \lambda}, \tau) $ is reflected in the choice of 
stochastic gauges.  Clearly, the aim is to find a gauge which will suppress  boundary terms which 
could  potentially show up in the partial integration step required to obtain the  Fokker-Planck 
equation. Here, we introduce drift 
gauges and then propose some first ideas on how to choose
the appropriate gauge. 

To formulate stochastic gauge invariance, it is useful to introduce the index 
$ \mu : 0 \cdots N_s^2$  such that  for example $ \underline{\pmb \lambda}_{\mu=0} =  \Omega $ and 
$ \underline{\pmb \lambda}_{\mu} = {\pmb n}_{x_{\mu},y_{\mu}} $ 
for $ \mu : 1 \cdots N_s^2$.  Then Eq.~(\ref{Anticomm}) 
may conveniently be written as: 
\begin{eqnarray}
- \frac{1}{2} \left[ \hat{H}, \hat{\Lambda}(\underline{\pmb \lambda})  \right]_{+}   & = &  
        \left(      
       - \sum_{\mu} \underline{{\pmb A}} _{\mu} \frac{\partial}
                              {\partial \underline{\pmb \lambda}_\mu } \right.     \\
 & + &\frac{1}{2} \sum_{\vec{i}, \mu,\nu } \underline{\pmb B}^{(\vec{i})}_{\mu} 
      \underline{\pmb B}^{(\vec{i})}_{\nu}  
\frac{\partial^2}{\partial \underline{\pmb \lambda}_\mu  
      \partial \underline{\pmb \lambda}_{\nu} }   \nonumber \\
 & +  &     \left.
\frac{1}{2} \sum_{\vec{i}, \mu, \nu } \underline{\pmb C}^{(\vec{i})}_{\mu} 
     \underline{\pmb C}^{(\vec{i})}_{\nu}  
\frac{\partial^2}{\partial \underline{\pmb \lambda}_{\mu} \partial \underline{\pmb \lambda}_{\nu} }
 \right)  
  \hat{\Lambda}(\underline{\pmb \lambda}), \nonumber 
\end{eqnarray}
with $ \underline{\pmb A} = (\Omega h({\pmb n}), {\pmb A} ) $,
$ \underline{\pmb B}^{(\vec{i})} = (0, {\pmb B}^{(\vec{i})} ) $ and 
with $ \underline{\pmb C}^{(\vec{i})} = (0, {\pmb C}^{(\vec{i})} ) $.
The above equation remains invariant under the transformation: 
\begin{eqnarray}
        & & \underline{\pmb B}^{(\vec{i})} = (0, {\pmb B}^{(\vec{i})} ) \rightarrow 
         (\Omega g^{(\vec{i})}, {\pmb B}^{(\vec{i})} ),   \\
        & & \underline{\pmb C}^{(\vec{i})} = (0, {\pmb C}^{(\vec{i})} ) \rightarrow 
         (\Omega f^{(\vec{i})}, {\pmb C}^{(\vec{i})} ), \nonumber \\ 
        & & \underline{\pmb A} = (\Omega h({\pmb n}), {\pmb A} ) \rightarrow
        (\Omega h({\pmb n}), {\pmb A} + \sum_{\vec{i}} g^{(\vec{i})} {\pmb B}^{(\vec{i})} 
              +f^{(\vec{i})} {\pmb C}^{(\vec{i})} ),  \nonumber 
\end{eqnarray}
where $g^{(\vec{i})}$ and $f^{(\vec{i})}$ are arbitrary functions of ${\pmb n }$.
This invariance stems from the fact that 
\begin{equation}
        \Omega \frac{\partial}{\partial \Omega} \hat{\Lambda}(\underline{\pmb \lambda}) = 
\hat{\Lambda}(\underline{\pmb \lambda}).
\end{equation}
For a given stochastic gauge  the Ito SDE reads: 
\begin{eqnarray}
\label{SDE_st}
 {\rm d } {\Omega}  &  = & - \Omega \left[ h({\pmb n}) {\rm d} \tau   
             - \sum_{\vec{i}} g^{(\vec{i})} {\rm d} W_{\vec{i}}   
             - \sum_{\vec{i}} f^{(\vec{i})} {\rm d} W_{\vec{i}}' \right] ,  \nonumber  \\
 {\rm d } {\pmb n} & = & - \underbrace{\left[ {\pmb A } + \sum_{\vec{i}} g^{(\vec{i})} {\pmb B}^{(\vec{i})} 
              + f^{(\vec{i})} {\pmb C}^{(\vec{i})} \right]}_{= \tilde{\pmb A} } {\rm d} \tau    \\ 
   & &    +  \sum_{\vec{i}} {\pmb B}^{(\vec{i})} {\rm d} W_{\vec{i}} + 
   \sum_{\vec{i}} {\pmb C}^{(\vec{i})} {\rm d } W'_{\vec{i}}. \nonumber
\end{eqnarray}

As apparent, one can modify the drift force  from ${\pmb A} $ to $\tilde{\pmb A} $ at 
the expense of adding noise in the weights  $\Omega$.  
Since our aim is to suppress the potential occurrence of boundary terms, 
one can follow the idea of choosing the gauge such that $\tilde{\pmb A} $ 
prohibits the walkers, $ {\pmb n} $,  of drifting to infinity. In other words 
$\tilde{\pmb A}_{x,y} $ should have the same sign as ${\pmb n}_{x,y} $.  
Fulfilling this requirement for each pair of indices $x,y$ leads to 
$ N_s^{2} $ equalities. But since we only have have  $2 N$  ( $g^{\vec{i}}$ and  $f^{\vec{i}}$ for 
$\vec{i} : 1 \cdots N $ )  degrees of 
freedom we can only fulfill the above condition on average. Defining 
a scalar product: 
\begin{equation}
        \left( {\pmb n} , \tilde{\pmb A }  \right)  \equiv  
\sum_{x,y} {\pmb n}_{x,y} \tilde{\pmb A}_{x,y}
\end{equation}
we require that 
\begin{equation}
        \left( {\pmb n} , \tilde{\pmb A}  \right)  > 0. 
\end{equation}

\section{Diffusion gauges -- positive-$P$ representation for the general electronic structure problem}
\label{diffusion_gauge}

One can also take advantage of gauge degrees of freedom by adding terms to the Hamiltonian which cancel each other (or are identically zero) and hence do not affect the observables. If these terms are of fourth and second order in the fermionic operators, they add a contribution to the diffusion part of the SDE, which is compensated in the drift part, while the equation for the weight $\Omega$ remains unaffected. 
In this appendix we will show how such a diffusion gauge allows one to generalize the positive-$P$ representation of the fermionic Hubbard model to the general electronic structure problem, including arbitrary hybridization, Coulomb and exchange terms. 

Since electronic structure calculations have important applications in quantum chemistry and the corresponding Hamiltonian
\begin{eqnarray}
\hat H &=& -\sum_{i\ne j\sigma} t_{ij} \hat c^\dagger_{i\sigma}\hat c_{j\sigma}
-\mu\sum_{i\sigma} \hat c^\dagger_{i\sigma} \hat c_{i\sigma}\nonumber\\
&&+\frac{1}{2}\sum_{ijkl\sigma\sigma'}V_{ijkl}\hat c^\dagger_{i\sigma} \hat c^\dagger_{k\sigma'} \hat c_{l\sigma'} \hat c_{j\sigma} \label{qchem}
\end{eqnarray}
has even been dubbed the ``theory of everything'' \cite{Laughlin}, a simulation approach without uncontrolled approximations or systematic errors is highly desired. In 
Eq.~(\ref{qchem}), $\hat c_{i\sigma}$ denotes the destruction operator for an electron with spin $\sigma$ in the orbital $i$, $t_{ij}$ the hopping amplitude and $\mu$ the chemical potential. The four-fermion terms $\hat c^\dagger_{i\sigma} \hat c^\dagger_{k\sigma'} \hat c_{l\sigma'} \hat c_{j\sigma}$ arise from Coulomb interactions.

The many-body problem (\ref{qchem}) can be mapped to a system of stochastic differential equations with positive weights, by an appropriate choice of diffusion gauge, which generalizes the gauge proposed by Corney and Drummond for the Hubbard model \cite{Corney04, C&D2}. 
Similar to section II, we choose a basis of Gaussian operators parameterized by $\underline{\pmb \lambda}$, which can be represented in this case by block-diagonal matrices with $N_o^2$ elements for spin up and spin down, respectively ($N_o$ is the number of orbitals). The Fokker-Planck equation \eqref{Fokkerplanck} for the probability
distribution $P(\underline{\pmb \lambda},\tau)$ reads
\begin{equation}
\frac{d}{d\tau}P(\underline{\pmb \lambda},\tau) = L[P(\underline{\pmb \lambda}, \tau)].\label{Fokker_Planck_precursor} 
\end{equation}
For a suitably chosen diffusion gauge, the operator $L$ in Eq.~(\ref{Fokker_Planck_precursor}) 
takes the form
\begin{align}
L = -\sum_\alpha \frac{\partial}{\partial \lambda_\alpha} A_\alpha &+ \frac{1}{2}\sum_{\alpha\beta}\frac{\partial}{\partial \lambda_\alpha} B_{\alpha} \frac{\partial}{\partial \lambda_\beta} B_{\beta}\nonumber\\
&+\frac{1}{2}\sum_{\alpha\beta}\frac{\partial}{\partial \lambda_\alpha} C_{\alpha} \frac{\partial}{\partial \lambda_\beta} C_{\beta}
\label{L}
\end{align}
with $B_\alpha$ and $C_\alpha$ real coefficients (we present here the formulation which leads to the smallest number of noise terms).
In this case, the Monte Carlo sampling can be done by integrating the Stratonovich SDE \cite{Gardiner}
\begin{equation}
d \lambda_\alpha(\tau) = A_\alpha(\underline{\pmb \lambda})d\tau + B_{\alpha}(\underline{\pmb \lambda}) dW(\tau)+ C_{\alpha}(\underline{\pmb \lambda}) dW'(\tau),\label{Stratonovich}
\end{equation}
where $\langle dW(\tau) dW'(\tau)\rangle=0$ and $\langle dW(\tau)^2\rangle=\langle dW'(\tau)^2\rangle=d\tau$. 

Gauge degrees of freedom can be used to modify the form of the operator $L$ in (\ref{Fokker_Planck_precursor}). 
In Ref.~\cite{Corney04}, the identity 
\begin{equation}
\hat n_{ii\sigma}^2 - \hat n_{ii\sigma} = 0
\end{equation}
was used to map the Hubbard model to a system of real SDE. Here we will show how the identity 
\begin{equation}
{\hat n_{ij\sigma}}^2-\delta_{ij}\hat n_{ij\sigma}=0\label{gauge}
\end{equation}
can be used to obtain real SDE and positive weights $\Omega$
for the more general Hamiltonian (\ref{qchem}). First, we note that the latter can be written as
\begin{equation}
\hat H = -\sum_{ij\sigma} t_{ij} \hat n_{ij\sigma}+\sum_{ijkl\sigma\sigma'}W_{ijkl}\hat n_{ij\sigma}\hat n_{kl\sigma'},\label{qchem_w}
\end{equation}
with $[\hat n_{ij\sigma}, \hat n_{kl\sigma'}]=0$. In particular, for $\sigma=\sigma'$ only terms with $i\ne k,l$ and $j\ne k,l$ appear. The $t_{ii\sigma}$ correspond to the chemical potential. We define
\begin{equation}
\hat H_{ij\sigma} = -t_{ij\sigma}\hat n_{ij\sigma},\label{H_t}
\end{equation}
and using Eq.~(\ref{gauge}) express the $ijkl\sigma\sigma'$-contribution in Eq.~(\ref{qchem}) in the form ($s_\W$ denotes the sign of $\W$)
\begin{eqnarray}
\hat H_\i &\equiv& \W\hat n_{ij\sigma}\hat n_{kl\sigma'}\nonumber\\
&=& -\frac{|\W|}{2}(\hat n_{ij\sigma}-s_\W\hat n_{kl\sigma'})^2\nonumber\\
&&+\frac{|\W|}{2}(\delta_{ij}\hat n_{ij\sigma}+\delta_{kl}\hat n_{kl\sigma'}).\label{H_W}
\end{eqnarray}
Each term $\hat H_m$ ($m=ij\sigma$ or $\i$) gives a contribution $A^{\Omega(n_{xy\rho})}_m$ to the drift term and the contributions $B^{n_{xy\rho}}_m$,  $C^{n_{xy\rho}}_m$ to the diffusion term of the (stochastic) differential equations (\ref{Stratonovich}) for $\Omega$ and $n_{xy\rho}$. No diffusion term appears in the equation of motion for $\Omega$ and we can write
\begin{eqnarray}
\frac{d\Omega}{d\tau} &=& \sum_m A^\Omega_m\Omega,\label{evolution_omega}\\
\frac{n_{xy\rho}}{d\tau} &=& \sum_m \Big( A^{n_{xy\rho}}_m+B^{n_{xy\rho}}_m\xi_m
+C^{n_{xy\rho}}_m\xi'_m\Big),\label{evolution_n}\hspace{3mm}
\end{eqnarray}
where the $\xi_m$, $\xi'_m$ are independent Gaussian random variables with variance $1/d\tau$.
The hopping term (\ref{H_t}) yields only the drift-contributions
\begin{eqnarray}
A^\Omega_{{ij\sigma}} &=& t_{ij\sigma}n_{ij\sigma},\label{A_2body}\\
A^{n_{xy\rho}}_{{ij\sigma}} &=& \frac{t_{ij\sigma}}{2}\big[n_{xj\sigma}(\delta_{iy}-n_{iy\sigma})+(\delta_{xj}-n_{xj\sigma})n_{iy\sigma}\big]\delta_{\rho\sigma}.\nonumber\\
\end{eqnarray}
With the gauge choice (\ref{H_W}), the interaction terms yield a Fokker-Planck equation of the form of Eqs.~(\ref{Fokker_Planck_precursor}) and (\ref{L}) with drift terms 
\begin{eqnarray}
A^\Omega_\i &=& -\W n_{ij\sigma}n_{kl\sigma'}+
\W n_{il\sigma}n_{kj\sigma'}\delta_{\sigma\sigma'},\label{A_4body}\\
A^{n_{xy\rho}}_\i &=& \frac{|\W|}{2}\Big\{[n_{xj\sigma}(\delta_{yi}-n_{iy\sigma})+(\delta_{xj}-n_{xj\sigma})n_{iy\sigma}]\nonumber\\
&&\hspace{3mm}\times (n_{ij\sigma}-s n_{kl\sigma'}-\delta_{ij}/2)\delta_{\rho\sigma}\nonumber\\
&&\hspace{0mm}+[n_{xl\sigma'}(\delta_{ky}-n_{ky\sigma'})+(\delta_{xl}-n_{xl\sigma'})n_{ky\sigma'}]\nonumber\\
&&\hspace{3mm}\times (n_{kl\sigma'}-s n_{ij\sigma}-\delta_{kl}/2)\delta_{\rho\sigma'}\Big\},
\end{eqnarray}
and diffusion terms
\begin{eqnarray}
B^{n_{xy\rho}}_\i &=& \sqrt{|\W|/2}\Big[n_{xj\sigma}(\delta_{iy}-n_{iy\sigma})\delta_{\rho\sigma}\nonumber\\
&&-s_\W n_{xl\sigma'}(\delta_{ky}-n_{ky\sigma'})\delta_{\rho\sigma'}\Big],\\
C^{n_{xy\rho}}_\i&=& \sqrt{|\W|/2}\Big[(\delta_{xj}-n_{xj\sigma})n_{iy\sigma}\delta_{\rho\sigma}\nonumber\\
&&-s_\W (\delta_{xl}-n_{xl\sigma'})n_{ky\sigma'}\delta_{\rho\sigma'}\Big].
\end{eqnarray}
Note that the right hand side of Eq.~(\ref{evolution_omega}) is $-h_m(\pmb n)\Omega$, with $h_m(\pmb n) = \text{Tr}(\hat \Lambda(\pmb n)\hat H_m)$. 
Furthermore, since the $n_{ij\sigma}$ are real variables, which remain real during the integration of Eq.~(\ref{evolution_n}), it follows from Eqs.~(\ref{A_2body}), (\ref{A_4body}) and (\ref{evolution_omega}) that the weight $\Omega$ will always stay positive.

For the actual implementation, it is simpler to use the Ito SDE, which is also numerically more stable.
In this case, the drift terms have to be modified as \cite{Gardiner}
\begin{align}
A_\alpha^\text{Ito} = A_\alpha^\text{Stratonovich}&+\frac{1}{2}\sum_{\beta}B_{\beta}\frac{\partial}{\partial \lambda_\beta}B_{\alpha}^{(n)}\nonumber\\
&+\frac{1}{2}\sum_{\beta}C_{\beta}\frac{\partial}{\partial \lambda_\beta}C_{\alpha}^{(n)},
\end{align}
while the diffusion terms remain unaffected. The four-fermion term (\ref{H_W}) thus yields a contribution
\begin{align}
&A^{n_{xy\rho}, \text{Ito}}_\i = -\frac{\W}{2}
\Big\{[n_{xj\sigma}(\delta_{iy}-n_{iy\sigma})\nonumber\\
&\hspace{3mm}+(\delta_{xj}-n_{xj\sigma})n_{iy\sigma}]n_{kl\sigma'}\delta_{\rho\sigma}\nonumber\\
&\hspace{0mm}+[n_{xl\sigma'}(\delta_{ky}-n_{ky\sigma'})\nonumber\\
&\hspace{3mm}+(\delta_{xl}-n_{xl\sigma'})n_{ky\sigma'}]n_{ij\sigma}\delta_{\rho\sigma'}\nonumber\\
&\hspace{0mm}-[(n_{xl\sigma}(\delta_{iy}-n_{iy\sigma})+(\delta_{xl}-n_{xl\sigma})n_{iy\sigma})n_{kj\sigma}\nonumber\\
&\hspace{3mm}+(n_{xj\sigma}(\delta_{ky}-n_{ky\sigma})+(\delta_{xj}-n_{xj\sigma})n_{ky\sigma})n_{il\sigma}]\delta_{\sigma\sigma'}\delta_{\rho\sigma}\Big\}
\label{sde_ito}
\end{align}
to the right hand side of (\ref{evolution_n}).

We have tested the Ito-SDE for the two-site model
\begin{eqnarray}
H&=&H_t+H_\mu+H_u+H_c+H_x+H_s+H_{h_1}+H_{h_2},\nonumber\\
\end{eqnarray}
where the individual terms are
\begin{eqnarray}
H_t &=& -t\sum_\sigma (\hat n_{12\sigma}+\hat n_{21\sigma}),\\
H_\mu &=& \mu\sum_\sigma (\hat n_{11\sigma}+\hat n_{22\sigma}),\\
H_u &=&  \frac{1}{2}\sum_\sigma (u_1\hat n_{11\sigma}\hat n_{11-\sigma}+u_2\hat n_{22\sigma}\hat n_{22-\sigma}),\\
H_c &=&  v_c\sum_\sigma (\hat n_{11\sigma}\hat n_{22\sigma}+\hat n_{11\sigma}\hat n_{22-\sigma}),\\
H_x &=&  v_x\sum_\sigma \hat n_{12\sigma}\hat n_{21-\sigma},\\
H_s &=&  v_s\sum_\sigma (\hat n_{12\sigma}\hat n_{12-\sigma}+\hat n_{21\sigma}\hat n_{21-\sigma}),\\
H_{h_1} &=&  h_1\sum_\sigma (\hat n_{11\sigma}\hat n_{12-\sigma}+\hat n_{11\sigma}\hat n_{21-\sigma}),\\
H_{h_2} &=&  h_2\sum_\sigma (\hat n_{22\sigma}\hat n_{12-\sigma}+\hat n_{22\sigma}\hat n_{21-\sigma}).
\end{eqnarray}
The (stochastic) differential equations (\ref{evolution_omega}) and (\ref{evolution_n}) were integrated using an implicit Euler scheme 
and adaptive time steps. Every $n_\text{reconf}$ steps, the family of random walkers was reconfigured according to the method of Ref.~\cite{Calandra98}. While the high-temperature behavior is correctly reproduced, similar and even more severe problems as those observed in the simulations of the Hubbard model -- most notably systematic errors in the energy at low temperature -- also plague the simulation of the two-site model with Coulomb interactions. We have not yet checked whether symmetry projections can be successfully applied in this case.

For realistic electronic structure calculations  
we propose to start with a Hartree Fock calculation of the electronic structure problem and to use both the occupied and unoccupied Hartree Fock orbitals in a subsequent quantum Monte Carlo simulation. This is the same Hamiltonian and basis set used in full-CI (configuration interaction) or coupled cluster methods (CCM) in quantum chemistry, but the algorithm described here would enable to study a larger number of basis functions than in a full-CI calculation. Using the Hartree Fock density matrix instead of a multiple of the unit matrix as initial density matrix $\hat{\rho}(0)$ has the advantage that the initial energy is already that of the Hartree-Fock solution, which will then be lowered further by the projection in imaginary time. Already a projection over a short imaginary time $\beta$ (which can now not be interpreted as inverse temperature) will give an energy lower than the Hartree-Fock solution.


\begin{thebibliography}{10}

\bibitem{Corney04}
J.~F. Corney and P.~D. Drummond, Phys. Rev. Lett. {\bf 93},  260401  (2004).

\bibitem{Zinn-Justin}
J. Zinn-Justin, {\em Quantum field theory and critical phenomena}, {\em The
  international series of monographs on physics.} (Clarendon Press, Oxford,
  1996).

\bibitem{Calandra98}
M.~C. Buonaura and S. Sorella, Phys. Rev. B {\bf 57},  11446  (1998).

\bibitem{Kloeden}
P. Kloeden, E. Platen, and H. Schurz, {\em Numerical solution of {S}{D}{E}
  through computer experiments} (Springer-Verlag, Berlin, 1994), with 1 IBM-PC
  floppy disk (3.5 inch; HD).

\bibitem{Mizusaki04}
T. Mizusaki and M. Imada, Phys. Rev. B {\bf 69},  125110  (2004).

\bibitem{Wigner}
E.~P. Wigner, {\em Group theory and its application to the quantum mechanics of
  atomic spectra}, {\em Pure and applied physics.} (Acad. Press, New York,
  1962).

\bibitem{Parola91}
A. Parola, S. Sorella, M. Parrinello, and E. Tosatti, Phys. Rev. B {\bf 43},
  6190  (1991).

\bibitem{Negele}
J.~W. Negele and H. Orland, {\em Quantum Many body systems}, {\em Frontiers in
  physics} (Addison-Wesley, Redwood City, Calif. u.a., 1988).

\bibitem{Laughlin}
R.~B. Laughlin and D. Pines, PNAS {\bf 97},  28  (2000).

\bibitem{C&D2}
J.~F. Corney and P.~D. Drummond, Cond-mat/0411712.

\bibitem{Gardiner}
C.~W. Gardiner, {\em Handbook of Stochastic Methods} (Springer,  1985).

\end{thebibliography}

\end{document}